\documentclass[pra,twocolumn,aps,showpacs,superscriptaddress]{revtex4}

\usepackage{amsmath}
\usepackage{amssymb}
\usepackage{hyperref}
\usepackage{graphicx}
\usepackage{subfigure}
\usepackage[usenames,dvipsnames]{color}
\usepackage{setspace}

\begin{document}

%%%%%%%%%%%%%%%%%%%%%%%%%%%%%%%%%%%%%%%%%%%%%%%%%%%%%%%%%%%%%%%%%%%%%%%%%%%%%%
%%%%                     Title and authors                                %%%%
%%%%%%%%%%%%%%%%%%%%%%%%%%%%%%%%%%%%%%%%%%%%%%%%%%%%%%%%%%%%%%%%%%%%%%%%%%%%%%

\title{Finite temperature dynamics of vortices in Bose-Einstein condensates}

\author{S. Gautam}
\affiliation{Department of Physics, Indian Institute of Science,
             Bangalore - 560 012, India}
\author{Arko Roy}
\affiliation{Physical Research Laboratory,
             Ahmedabad - 380 009, India}
\author{Subroto Mukerjee}
\affiliation{Department of Physics, Indian Institute of Science,
             Bangalore - 560 012, India}

%%%%%%%%%%%%%%%%%%%%%%%%%%%%%%%%%%%%%%%%%%%%%%%%%%%%%%%%%%%%%%%%%%%%%%%%%%%%%%
%%%%%%%%%%                    Abstract                             %%%%%%%%%%%
%%%%%%%%%%%%%%%%%%%%%%%%%%%%%%%%%%%%%%%%%%%%%%%%%%%%%%%%%%%%%%%%%%%%%%%%%%%%%%

\date{\today}
\begin{abstract}
We study the dynamics of a single and a pair of vortices in quasi
two-dimensional Bose-Einstein condensates at finite temperatures. We use the
stochastic Gross-Pitaevskii equation, which is the Langevin equation for
the Bose-Einstein condensate, to this end. For a pair of vortices, we study the
dynamics of both the vortex-vortex and vortex-antivortex pairs, which are
generated by rotating the trap and moving the Gaussian obstacle potential,
respectively. Due to thermal fluctuations, the constituent vortices are not
symmetrically generated with respect to each other at finite temperatures.
This initial asymmetry coupled with the presence of random thermal fluctuations
in the system can lead to different decay rates for the component vortices of
the pair, especially in the case of two corotating vortices.
\end{abstract}


\maketitle

%%%%%%%%%%%%%%%%%%%%%%%%%%%%%%%%%%%%%%%%%%%%%%%%%%%%%%%%%%%%%%%%%%%%%%%%%%%%%%
%%%%%%%%%%%%%                 Introduction                         %%%%%%%%%%%
%%%%%%%%%%%%%%%%%%%%%%%%%%%%%%%%%%%%%%%%%%%%%%%%%%%%%%%%%%%%%%%%%%%%%%%%%%%%%%

\section{Introduction}
\label{I}
In the spin zero or scalar Bose-Einstein condensates (BECs), vortices carry 
integral angular momentum and serve as the unambiguous proof of superfluidity 
of these systems. The various methods which have been employed by the 
experimentalists to generate the vortices in BECs include manipulating the 
interconversion between the internal spin states of an isotope \cite{Matthews},
stirring the BEC with the laser beam \cite{Madison, Raman}, rotating the BEC 
\cite{Abo-Shaeer}, phase imprinting \cite{Leanhardt}, etc. Vortex dipoles, 
consisting of vortex-antivortex pairs, have also been experimentally realized 
in BECs by moving the condensate across the Gaussian obstacle potential 
\cite{Neely}. Vortex dipoles are also formed as the decay product of the 
soliton in quasi two-dimensional condensates \cite{Anderson}. More recently, 
formation of the vortex dipoles during the rapid quench through the 
condensation temperature has also been observed \cite{Weiler,Freilich}
The dynamics of a single vortex and a vortex-antivortex pair has also been 
observed experimentally \cite{Freilich}. On theoretical front, the dynamics 
of vortex dipoles has also been analyzed at zero temperature 
\cite{Middelkamp,Torres}. Recently, the dynamics of small vortex clusters 
with $2$-$4$ corotating vortices was also studied both experimentally and 
theoretically \cite{Navarro}. The decay of an off-centered vortex at finite 
temperature has been studied using pure classical field treatment 
\cite{Schmidt}, dissipative Gross-Pitaevskii equation (DGPE) 
\cite{Madarassy}, Zaremba-Nikuni-Griffin (ZNG) formalism \cite{Jackson,Allen}, 
projected Gross-Pitaevskii equation (PGPE) \cite{Wright}, and stochastic 
projected Gross-Pitaevskii equation (SPGPE) \cite{Rooney}. 

To the best of our knowledge, the finite temperature dynamics of a pair of
vortices has not been investigated theoretically which we do in the present 
work, along with a study of the dynamics of a single vortex. The method we 
employ is the stochastic Gross-Pitaevskii equation (SGPE) 
\cite{Stoof-1,Duine-1,Stoof-2}. For the case of a single vortex in a 
condensate with Thomas-Fermi density profile, the present work is the first 
one to employ the SGPE, which is a better method than ZNG in low-dimensional 
systems near critical temperatures. In weakly interacting domain, where the 
condensate has the Gaussian density profile, the stochastic equations of 
motion for the position of vortex core have also been derived using a 
variational approach to the SGPE 
\cite{Duine-2}. 
The SGPE has been successfully used to study finite temperature
scalar BECs in quasi one- and two-dimensional geometries
\cite{Proukakis-1,Cockburn-1,Cockburn-2,Cockburn-3,Cockburn-4,Cockburn-5}.

The paper is organized as follows: In Sec.~\ref{II}, we describe the SGPE
method. We then discuss the finite temperature dynamics of a single vortex
in Sec.~\ref{III}. This is followed by the investigation of
the dynamics of a vortex pair in Sec.~\ref{IV}. We conclude by providing
the summary and conclusions in the last section.

%%%%%%%%%%%%%%%%%%%%%%%%%%%%%%%%%%%%%%%%%%%%%%%%%%%%%%%%%%%%%%%%%%%%%%%%%%%%%%
%%%%%%%%%%%%%                    SGPE                              %%%%%%%%%%%
%%%%%%%%%%%%%%%%%%%%%%%%%%%%%%%%%%%%%%%%%%%%%%%%%%%%%%%%%%%%%%%%%%%%%%%%%%%%%%

\section{Stochastic Langevin equation for the BEC}
\label{II}
At $T=0~$K, a scalar BEC is well described by the Gross-Pitaevskii (GP) equation
\cite{Dalfovo}
\begin{equation}
i\hbar\frac{\partial \Phi(\mathbf x,t)}{\partial t} =
\left(-\frac{\hbar^2\nabla^2}{2m}
+ V(\mathbf x) +g|\Phi(\mathbf x,t)|^2\right) \Phi(\mathbf x,t),
\label{Eq.1}
\end{equation}
where $\Phi(\mathbf x, t)$ is the wave function of the BEC.  The trapping
potential
$V (\mathbf x) = (\omega_x^2 x^2 + \omega_y^2 y^2 + \omega_z^2 z^2 )/(2m)$.
The interaction between atoms with mass $m$ is characterized by the
interaction strength $g = 4\pi\hbar^2 a/m$, where $a$ is the
$s$-wave scattering length. The GP equation conserves the total number of
atoms and total energy. In order to study the finite temperature scalar BEC,
we use the stochastic Gross-Pitaevskii equations (SGPE)
\cite{Cockburn-1}
\begin{eqnarray}
i\hbar\frac{\partial \Psi(\mathbf x,t)}{\partial t}
&=& (1-i\gamma)\left(-\frac{\hbar^2\nabla^2}{2m}
+ V +g|\Psi(\mathbf x,t)|^2-\mu\right)\nonumber\\
& & \Psi(\mathbf x,t)+ \eta(\mathbf x,t),
\label{Eq.2}
\end{eqnarray}
where $\gamma$ is the dissipation, $\mu$ is the chemical potential, and
$\eta(\mathbf x,t)$ is the random fluctuation which satisfies the following
noise-noise correlation
\begin{equation}
\langle \eta(\mathbf x,t) \eta^*(\mathbf x',t')\rangle = 2\gamma k_B T\hbar
                             \delta(\mathbf{ x-x'})\delta(\mathbf{t-t'}),
\end{equation}
according to the fluctuation-dissipation theorem required for equilibration,
where $k_B$ and $T$ are the Boltzmann constant and temperature respectively, 
and $\langle \ldots \rangle$ denotes the averaging over different noise
realizations. The system described by the SGPE is assumed to be divided into 
two parts: The first consisting of a few highly occupied low lying modes is
represented by a Langevin field $\Psi(\mathbf x,t)$ and the second,
representing a heat bath, is denoted by the noise $\eta$. The effect of the 
higher modes is taken into account by the noise. The Langevin field
is the expectation value of the annihilation field operator $\hat{\Psi}$ at 
zero temperature and at finite temperature has contributions from thermal 
fluctuations, i.e.,
\begin{equation}
\Psi = \langle\hat{\Psi}\rangle + \delta\hat{\Psi}
                        \equiv \Phi + \delta \hat{\Psi},
\end{equation}
where $\Phi$ is the solution of Eq.~(\ref{Eq.1}). While numerically solving 
the SGPE, the division into the low lying modes represented by the Langevin 
field and all the higher modes represented by the noise is automatically
implemented by numerical discretisation. The discretisation introduces the 
ultraviolet momentum cutoff, which separates the multimode Langevin field from 
the higher modes. An important assumption that is made while deriving the SGPE 
equation is that the thermal atoms are assumed to have a Rayleigh-Jeans 
distribution with non-zero chemical potential instead of a Bose-Einstein 
distribution. Due to this assumption, the dynamics obtained from the SGPE is 
sensitive to the ultraviolet energy cutoff imposed on the system. In the 
numerical simulations, the energy cutoff and the finite grid spacing are 
related by
\begin{equation}
 E_{\rm cutoff} = \frac{\hbar^2|\mathbf k_{\rm cutoff}|^2}{2m}
                = \frac{\pi k_BT}{2} \left(\frac{\lambda_T}{\alpha}\right)^2,
\end{equation}
where $\alpha$ is the grid spacing in the square lattice and $\lambda_T$ is 
the thermal de Broglie wavelength. In the present work, we consider 
$\delta x=\delta y=\alpha=0.11$, where $\delta x$ and $\delta y$ are the
grid spacings along the $x$ and $y$ directions respectively. Before 
proceeding further, we transform the SGPE into dimensionless form using
following transformations:
\begin{eqnarray}
\mathbf x &= &\tilde{\mathbf {x}} a_{\rm osc},~t  =
\frac{2\tilde {t} \omega^{-1}}{1+\gamma^2},~
\Psi = \frac{\tilde {\Psi}}{a_{\rm osc}^{3/2}},~
\eta = \frac{\hbar\omega}{a_{\rm osc}^{3/2}} \tilde{\eta}
\end{eqnarray}
where $a_{\rm osc} = \sqrt{\hbar/(m\omega)}$ with $\omega =
\rm min~ \{\omega_x,\omega_y,\omega_z\}$ is the oscillator length.
The dimensionless SGPE equation for the BEC is now of the form

\begin{eqnarray}
(i-\gamma)\frac{\partial \tilde{\Psi}}{\partial \tilde{t}}
&=& \left(-\tilde{\nabla}^2 + 2\tilde{V} +2\tilde{g}|\tilde{\Psi}|^2
-2\tilde{\mu}\right) \tilde{\Psi}\nonumber\\
& &+2\frac{\tilde{\eta}(\tilde {\mathbf x},\tilde{t})}{1-i\gamma},
\label{Eq.4}
\end{eqnarray}
where $\tilde V = (\lambda_x^2\tilde{x}^2+\lambda_y^2\tilde{y}^2
+\lambda_z^2\tilde{z}^2)/2$. Also, $\lambda_\alpha = \omega_\alpha/\omega$
with $\alpha = x,y,z$ and $\tilde g = 4\pi\hbar Na/(3m\omega a_{\rm osc}^3)$.
The noise-noise correlation in scaled coordinates now reads 
\begin{equation}
\langle \tilde{\eta}(\tilde {\mathbf x},\tilde{t}) 
\tilde{\eta}^*(\tilde{\mathbf x}',
\tilde{t}')\rangle = \gamma \frac{k_B T}{\hbar\omega} (1+\gamma^2)
\delta(\tilde{\mathbf x}-\tilde {\mathbf x}')\delta(\tilde{t}-\tilde{t}').
\end{equation}
For the sake of simplicity of notations, from here on we will show the scaled
variables without tildes in the rest of the manuscript except if mentioned
otherwise. In the present work, we consider the BEC in quasi two-dimensional
traps for which $\lambda_x=\lambda_y=1\ll\lambda_z$. In this case, the axial
degrees of freedom of the system are frozen. We write
$\Psi(\mathbf x) = \zeta(z)\psi(x,y)$ with
$\zeta(z) = (\lambda_z/\pi)^{1/4}e^{-(\lambda_z z^2)/2}$ as the harmonic
oscillator ground state along the axial direction. This allows us to use the
following two-dimensional equation after integrating out the axial coordinate:
\begin{eqnarray}
(i-\gamma)\frac{\partial \psi}{\partial t} &=& \left(-\nabla_{xy}^2
+ 2V_{xy} +2g_{xy}|\psi|^2-2\mu\right) \psi\nonumber\\
 & &+\frac{2\eta}{1-i\gamma},
\label{sgpe_scaled}
\end{eqnarray}
where $\nabla_{xy}^2 = \partial^2/\partial x^2+ \partial^2/\partial y^2$,
$V_{xy} = x^2/2+y^2/2$ and $g_{xy} = \sqrt{\lambda_z/2\pi}g$. We use time
splitting Fourier spectral method to solve equation Eq.~(\ref{sgpe_scaled})
on $256\times256$ square grid \cite{Bao}. The spatial and time step sizes
employed in the present work are $0.11$ and $0.006$ respectively.

In the SGPE, physical properties are calculated by averaging over several
independent noise realizations. Thus, it has a close resemblance to
experiments, where relevant data is obtained by averaging over several
independent experimental realizations. In other words, the individual results
obtained with independent noise realizations are equivalent to the individual
results obtained from independent experimental realizations. Due to the random
nature of the noise, the results of various noise realizations will differ
from one another as is the case in experiments. Herein lies the first advantage
of SGPE method as it can account for experimental shot to shot variation in
the properties of the system, whereas methods like the ZNG can very accurately
account only for the average properties. The second advantage is that in 
contrast to the ZNG formalism, the SGPE method can also fully describe the 
regimes where the thermal fluctuations in the condensate phase are quite large,
e.g. one- and two-dimensional systems at temperatures close to the transition 
temperature. Among the other methods used to study finite temperature BECs,
the stochastic projected Gross-Pitaevskii equation (SPGPE) has a resemblance 
to the SGPE \cite{Blakie,Gardiner-1,Gardiner-2}. The crucial difference is the 
use of projector to extract the low lying coherent modes in SPGPE.

The SGPE simulates the system in a grand canonical ensemble since there is no 
conservation of the atom number. The equilibrium value of the number of atoms 
is fixed by the chemical potential. While studying the dynamics, especially 
in the presence of strong perturbations the fluctuations in atom number above 
the mean may be quite large. However, at very low temperatures or over small 
time scales after the system has reached equilibrium, the atom number may be 
approximately conserved. This situation can be mimicked by first generating 
the equilibrium solution using the SGPE and then switching of the dissipation 
and noise, i.e., setting $\gamma=\eta=0$ to study the dynamics. This method 
ensures atom number conservation over the regime of interest and is denoted 
as SGPE$_{\rm eq}$ in this paper. This method has been earlier used to model 
the quasi condensate growth on an atom chip \cite{Proukakis-2} and is similar 
in spirit, to the truncated Wigner method \cite{Steel}.

%%%%%%%%%%%%%%%%%%%%%%%%%%%%%%%%%%%%%%%%%%%%%%%%%%%%%%%%%%%%%%%%%%%%%%%%%%%%%%
%%%%%%%%%%%%%           Dynamics of a single vortex                %%%%%%%%%%%
%%%%%%%%%%%%%%%%%%%%%%%%%%%%%%%%%%%%%%%%%%%%%%%%%%%%%%%%%%%%%%%%%%%%%%%%%%%%%%

\section{Dynamics of a single vortex}
\label{III}
We first study the dynamics of a single vortex, which is created in the BEC by
rotating it at an optimum frequency, at finite temperature. In the present work,
we consider $\approx1\times10^5$-$1.5\times10^5$ atoms of $^{87}$Rb trapped in
a trapping potential with $\omega = \omega_x=\omega_y=2\pi\times10$ Hz and
$\omega_z=2\pi\times100$ Hz. With this choice of parameters, 
$a_{\rm osc} = 3.41~\mu$m and $\omega^{-1} = 1.59\times10^{-2}$s, the respective
units of length and time employed in the manuscript.
The $s$-wave scattering length of $^{87}$Rb is $99a_0$.
Let us term this set of parameters as {\em parameters set a}. The critical 
temperature obtained by using the expression for an ideal three-dimensional 
Bose gas  
$T^{3d}_c = \hbar\bar{\omega}N^{1/3}/k_B$ is $\approx 45-52~{\rm nK},$
where $\bar{\omega} = (\omega_x \omega_y \omega_z)^{1/3}$. The expression is
valid provided $k_BT\gg \hbar\omega$. On the other hand, the transition 
temperature for a strictly two-dimensional ideal Bose gas is
 $T^{2d}_c = \sqrt{6N\omega_x \omega_y}\hbar/\pi k_B
            \approx 118-145 ~{\rm nK}.$ \cite{Bagnato}
In the quasi two-dimensional case with finite
$\tilde\omega_z = \hbar\omega_z/k_BT^{2d}_c$,
the critical temperature is significantly reduced \cite{Holzmann}, and for
the aforementioned parameters
$T^{q2d}_c = 0.37T^{2d}_c \approx 44-53  ~{\rm nK}$,
and hence is closer to the value for a three dimensional ideal Bose gas.
In the present work, we study the dynamics of vortices in BECs at
finite temperatures, which can be as high as $50$ nK, where thermal
fluctuations are appreciable.

The dissipation parameter $\gamma$ is in general a function of both space and 
time, but in the present work we consider it to be independent of spatial
and temporal coordinates. An approximate damping based on microscopic
considerations has been suggested by Penckwitt et al. \cite{Penckwitt} which 
gives us the damping parameter
\begin{equation}
\gamma = \kappa \frac{4mk_BT}{\pi}\left(\frac{a}{\hbar}\right)^2,
\label{diss_parameter}
\end{equation}
where $\kappa = 3$.
This choice of $\kappa = 3$ reproduces the growth rate observed in most
experiments \cite{Penckwitt}.

\begin{center}
\begin{figure}[!h]
\begin{tabular}{cc}
\resizebox{!}{!}
{\includegraphics[trim = 2cm 0mm 1.5cm 0mm,clip, angle=0,width=4.5cm]
                 {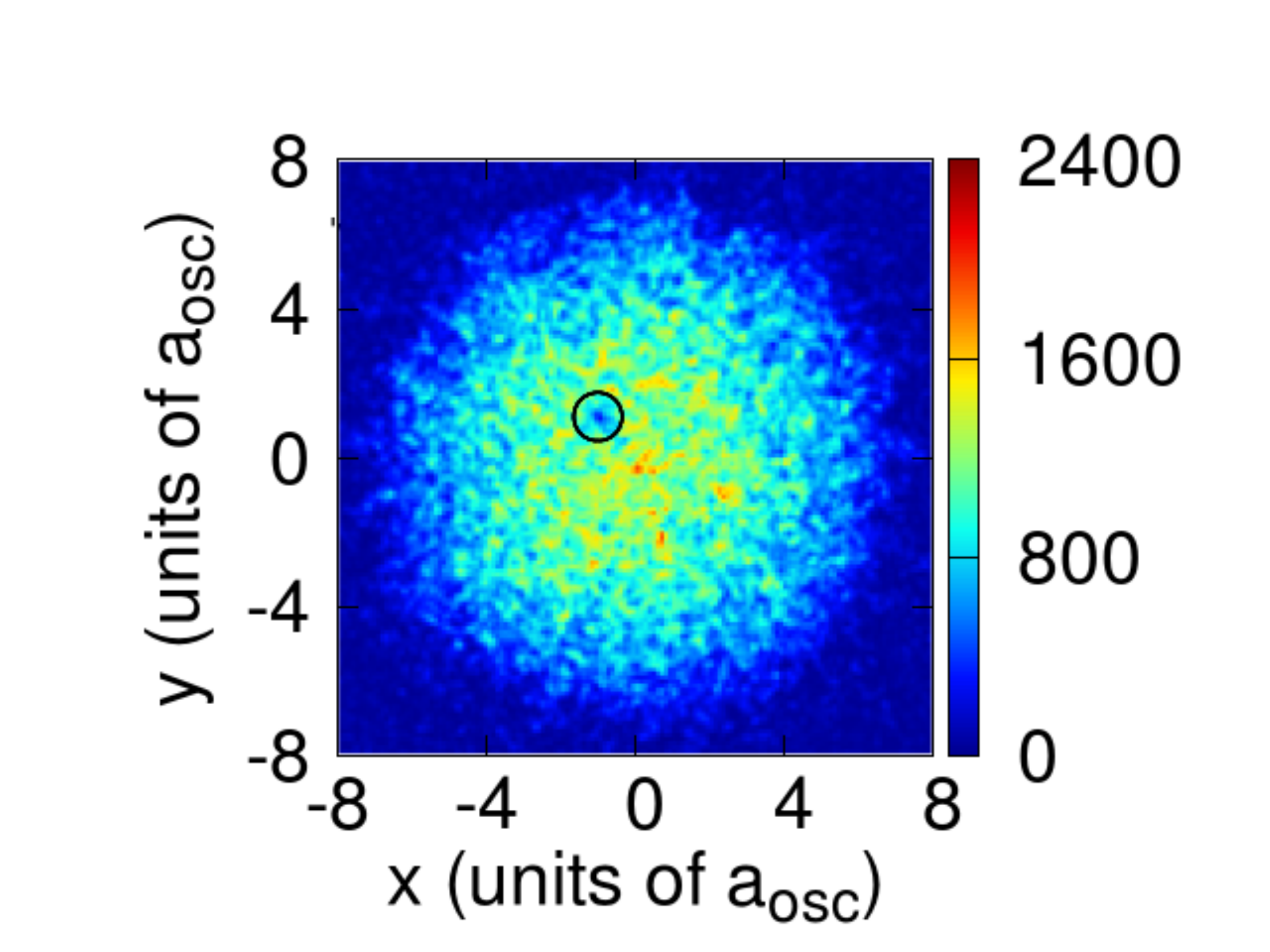}}
\resizebox{!}{!}
{\includegraphics[trim = 2cm 0mm 1.5cm 0mm,clip, angle=0,width=4.5cm]
                 {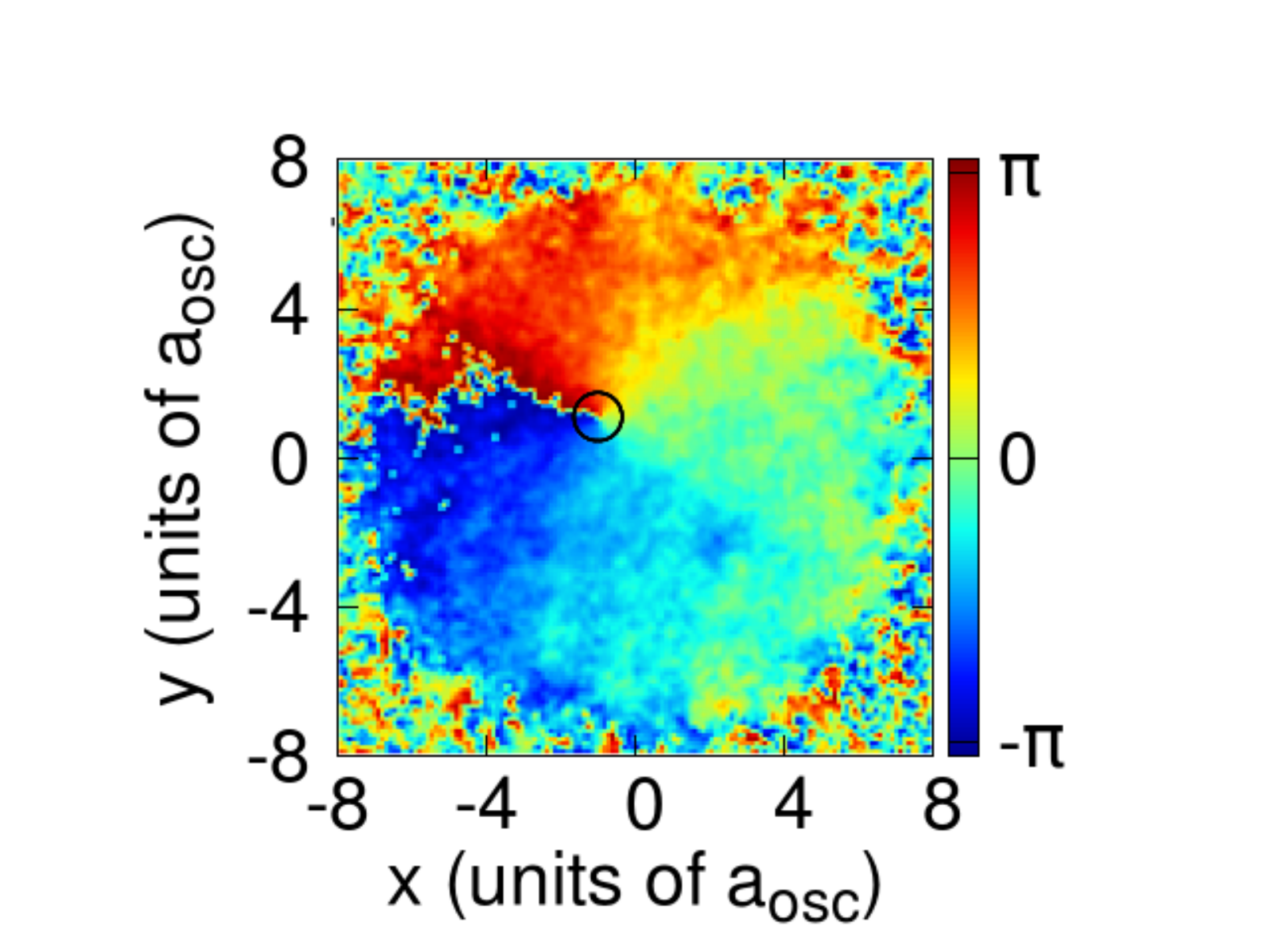}}\\
\end{tabular}
\includegraphics[trim = 0.5cm 0mm 0cm 0mm,clip, angle=0,width=8cm]
                 {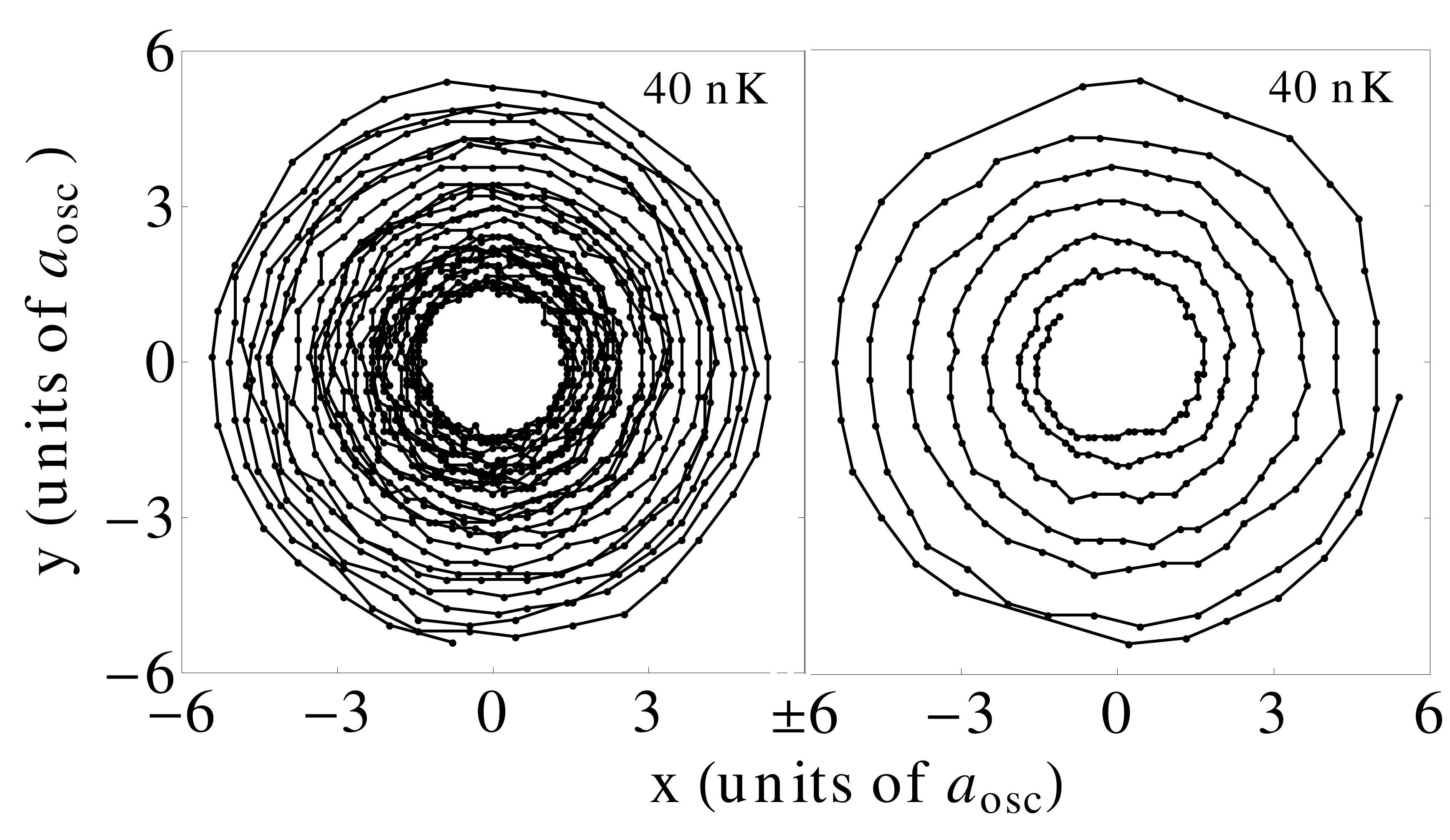}
\caption{(Color online) The upper panel shows the density (left image) and 
phase profile (right image) of the condensate just before switching off the 
rotational frequency at $T = 40$ nK. The location of the vortex has been 
marked by a black circle in both the density and phase profiles. The lower 
panel shows the trajectory traversed by the vortex for two different values of 
$\gamma$; $\gamma = 7.5\times10^{-4}$ (left) and 
$\gamma = 7.5\times10^{-3}$ (right).
}
\label{Fig-1}
\end{figure}
\end{center}

\begin{center}
\begin{figure}[!h]
\includegraphics[trim = 0.5cm 0mm 0cm 0mm,clip, angle=0,width=9cm]
                 {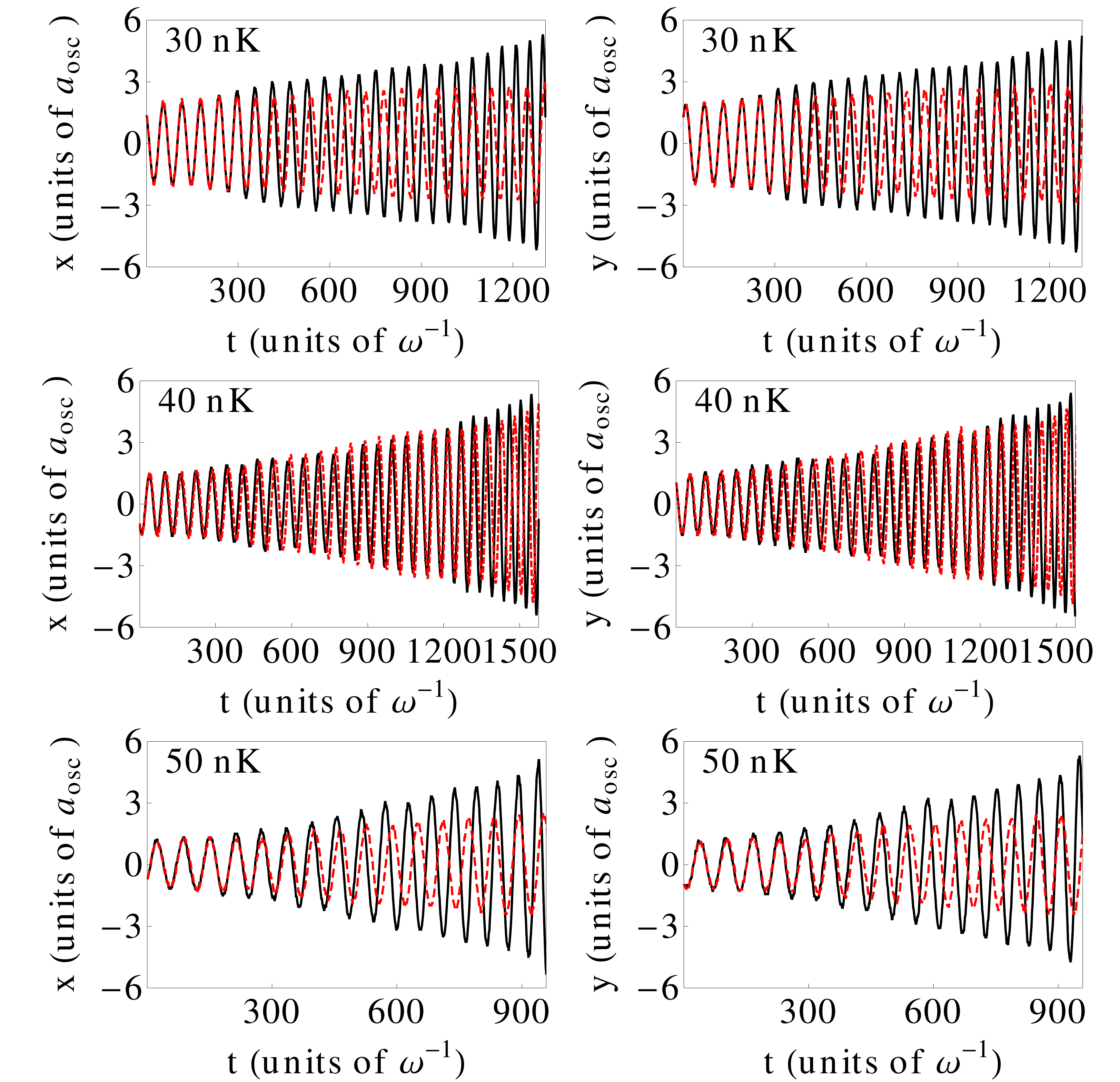}
\caption{(Color online) Dynamics of a single off-center vortex at $30$ nK 
(upper panel), $40$ nK (middle panel) and $50$ nK (lower panel).
The plots on the left show the $x$-coordinate of the vortex
as a function of time, while the plots on the right show the $y$-coordinate of
the vortex as a function of time. Solid-black and dashed-red curves
correspond to results obtained using SGPE and SGPE$_{\rm eq}$ respectively.
}
\label{Fig-2}
\end{figure}
\end{center}

\begin{center}
\begin{figure}[!h]
\includegraphics[trim = 0cm 0mm 0cm 0mm,clip, angle=0,width=6cm]
                 {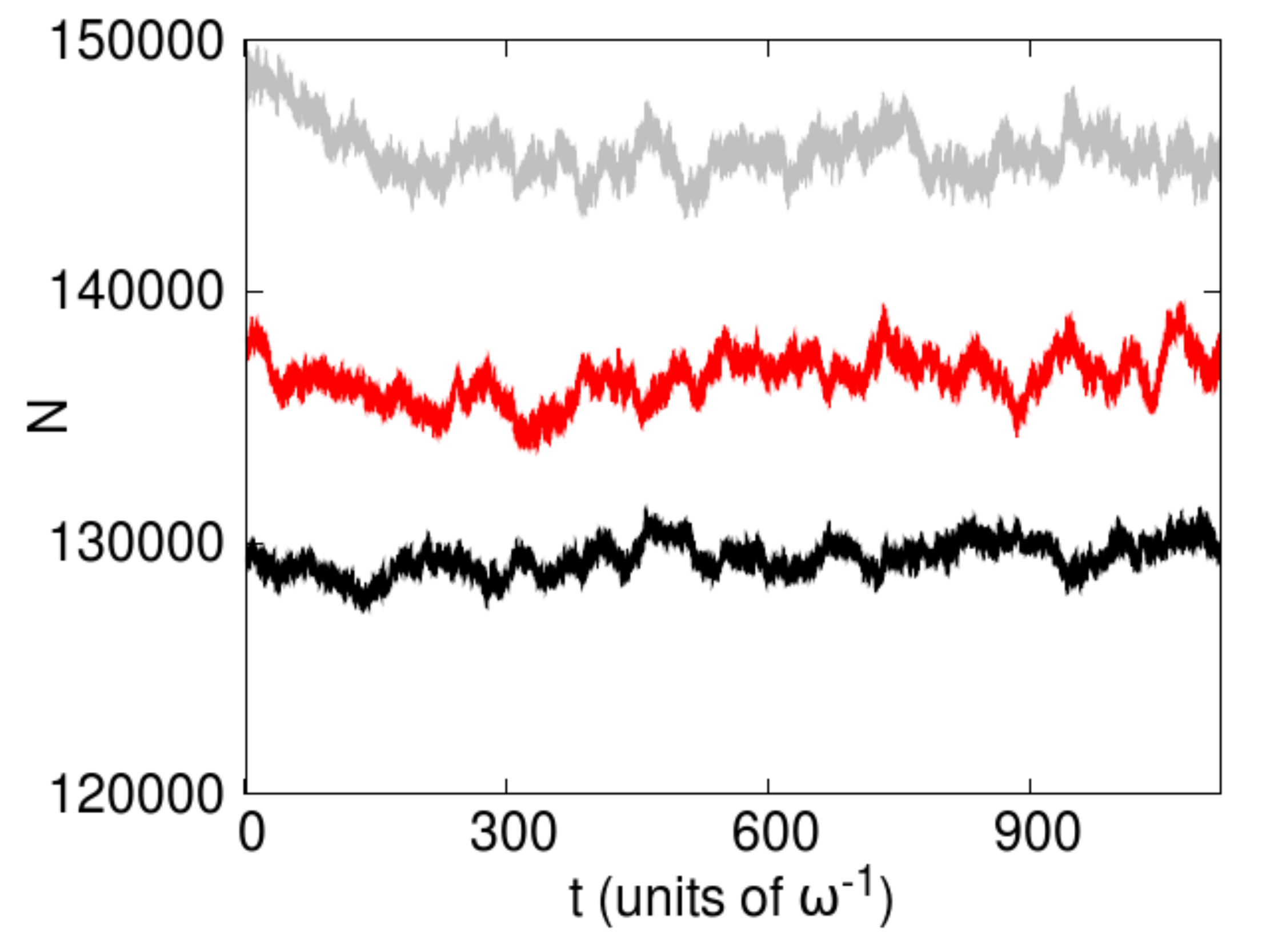}
\caption{(Color online) The plot shows the number of atoms as function of time,
obtained using SGPE, after switching off the $\Omega$. The black, red and grey 
curves correspond to $30$ nk, $40$ nK, and $50$ nK temperatures respectively. 
In SGPE$_{\rm eq}$, the number remains fixed to the value at $t=0$ at all
temperatures.}
\label{num_atoms1}
\end{figure}
\end{center}

We first generate a single vortex by rotating the condensate with a frequency
$\Omega = 0.1~\omega_x$ at $T=30$ nK, $40$ nK and $50$ nK.  Having generated a 
vortex, we switch off the rotation. The density and phase profiles of the 
condensate just prior to switching off the rotation can be seen in the upper 
panel of Fig.~\ref{Fig-1} for a temperature of $40$ nK.
In steady state, the position of the vortex is different at different 
temperatures. We let the system evolve after switching off the rotation and 
observe that with the passage of time the vortex slowly spirals out of the 
condensate for all the aforementioned temperatures. This leads to the decrease
in the energy of the system. As an example, the trajectories traversed by the 
vortex at $40$ nK for two different values of $\gamma$, 
$\gamma = 7.5\times10^{-4}$ obtained from Eq.~(\ref{diss_parameter}) and 
$\gamma = 7.5\times10^{-3}$, are shown in the lower panel of Fig.~\ref{Fig-1}.  
In both the cases, the vortex slowly spirals out of the condensate with a 
greater rate of decay for a larger value of $\gamma$. The exact dynamics of 
the off-center vortex at $T=30$ nK, $40$ nK and $50$ nK temperatures is shown 
in Fig.~\ref{Fig-2}.
We find that the results of number conserving SGPE$_{\rm eq}$ agree with those 
of the SGPE for the initial period of evolution. This is evident from the 
dynamics in Fig.~\ref{Fig-2} up to $t= 300~\omega^{-1}$. Within this period, 
the agreement between the SGPE$_{\rm eq}$ and the SGPE dynamics is better at 
lower temperatures. It implies that the effect of the noise, which accounts for
modes with energy higher than $E_{\rm cutoff}$, is negligible over small time 
scales. Over this period, the multimode Langevin field solely describes the 
system dynamics. The SGPE$_{\rm eq}$ results in much slower
decay of the vortex as compared to SGPE as is evidenced by the lower amplitude 
of oscillations of the $x$- and $y$-coordinates of the vortex in 
Fig.~\ref{Fig-2}. We find that the decay rate of the vortex depends on its 
initial location as well as temperature. In our simulations, the vortex 
created at $40$ nK takes a much longer time to decay than the vortex created 
at $30$ nK as is shown in Fig~\ref{Fig-2}. At $40$ nK the vortex, just prior 
to switching off the rotation, is located closer to trap center as compared to 
its location at $30$ nK. Hence, the density of the thermal atoms is lower in 
the vicinity of the vortex at $40$ nK than at the $30$ nK. This leads to the 
slower decay rate at $40$ nK temperature. These results are in qualitative 
agreement with previous studies \cite{Jackson, Allen, Rooney}. In the present 
work, as we cannot precisely control the initial location of the vortex, 
quantitative comparison with earlier studies is not possible. The actual 
variation in the number of atoms within the SGPE after the rotation is switched
off is shown in Fig.~\ref{num_atoms1}.

%%%%%%%%%%%%%%%%%%%%%%%%%%%%%%%%%%%%%%%%%%%%%%%%%%%%%%%%%%%%%%%%%%%%%%%%%%%%%%
%%%%%%%%%%%%%           Dynamics of a vortex pair                  %%%%%%%%%%%
%%%%%%%%%%%%%%%%%%%%%%%%%%%%%%%%%%%%%%%%%%%%%%%%%%%%%%%%%%%%%%%%%%%%%%%%%%%%%%

\section{Dynamics of a vortex pair}
\label{IV}
%%%%%%%%%%%%%%%%%%%%%%%%%%%%%%%%%%%%%%%%%%%%%%%%%%%%%%%%%%%%%%%%%%%%%%%%%%%%%%
%%%%%%%%%         Dynamics of two corotating vortices                %%%%%%%%%
%%%%%%%%%%%%%%%%%%%%%%%%%%%%%%%%%%%%%%%%%%%%%%%%%%%%%%%%%%%%%%%%%%%%%%%%%%%%%%

\subsection{Dynamics of two corotating vortices}
\begin{center}
\begin{figure}[h]
\begin{tabular}{cc}
\resizebox{!}{!}
{\includegraphics[trim = 2cm 2cm 1.5cm 0mm,clip, angle=0,width=4.4cm]
                 {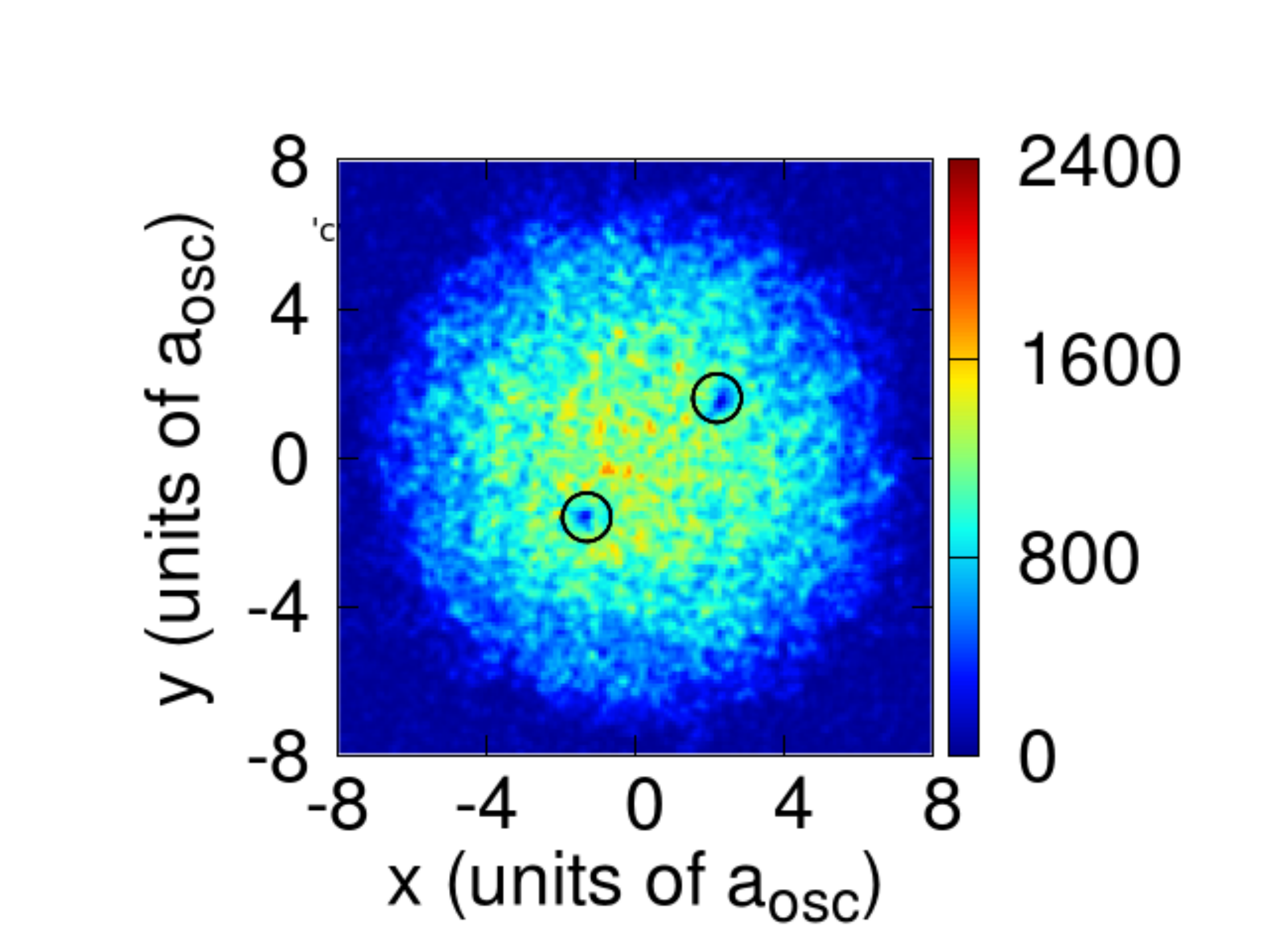}}
\resizebox{!}{!}
{\includegraphics[trim = 2cm 2cm 1.5cm 0mm,clip, angle=0,width=4.4cm]
                 {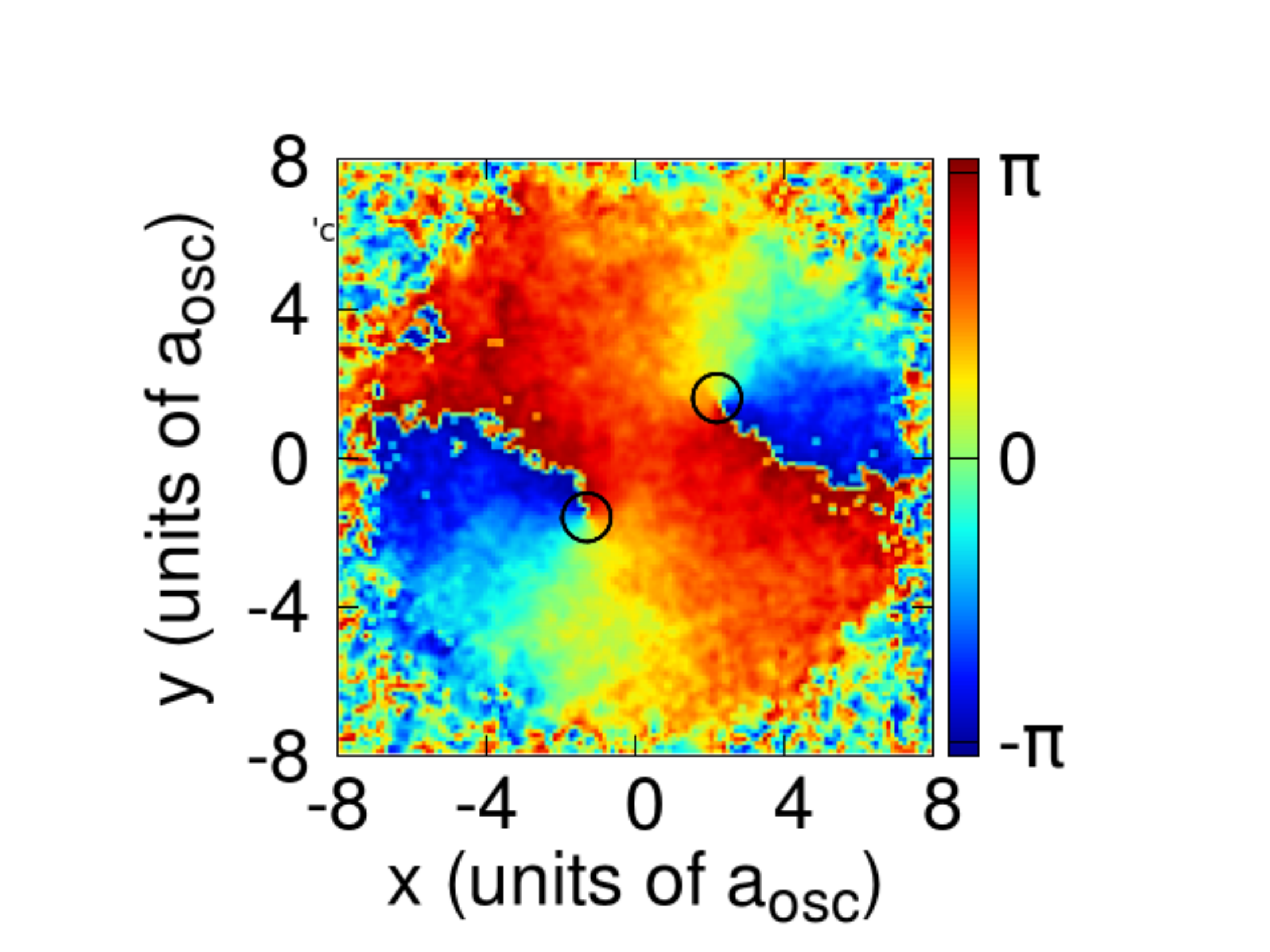}}\\
\resizebox{!}{!}
{\includegraphics[trim = 2cm 0mm 1.5cm 2cm,clip, angle=0,width=4.4cm]
                 {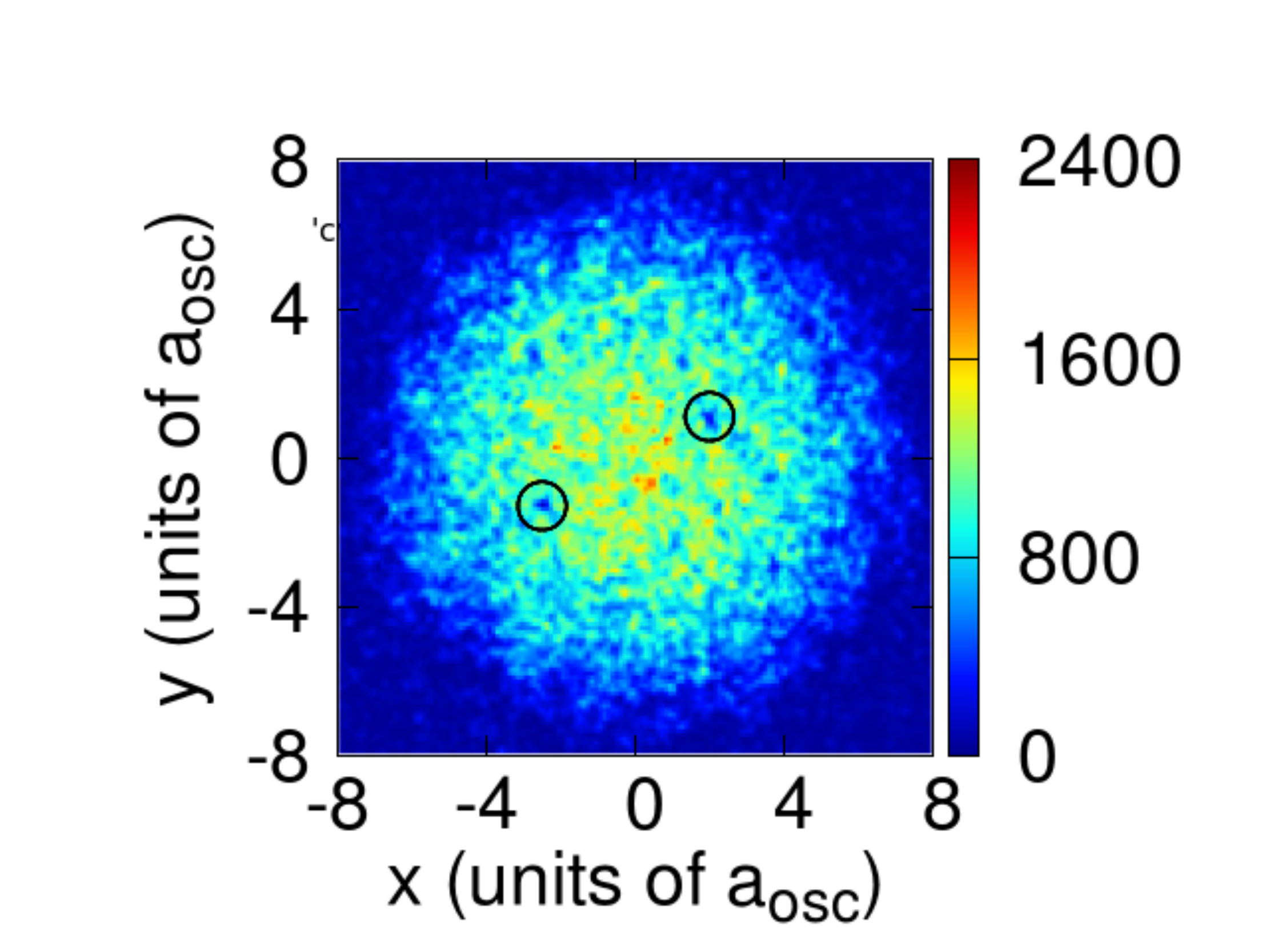}}
\resizebox{!}{!}
{\includegraphics[trim = 2cm 0mm 1.5cm 2cm,clip, angle=0,width=4.4cm]
                 {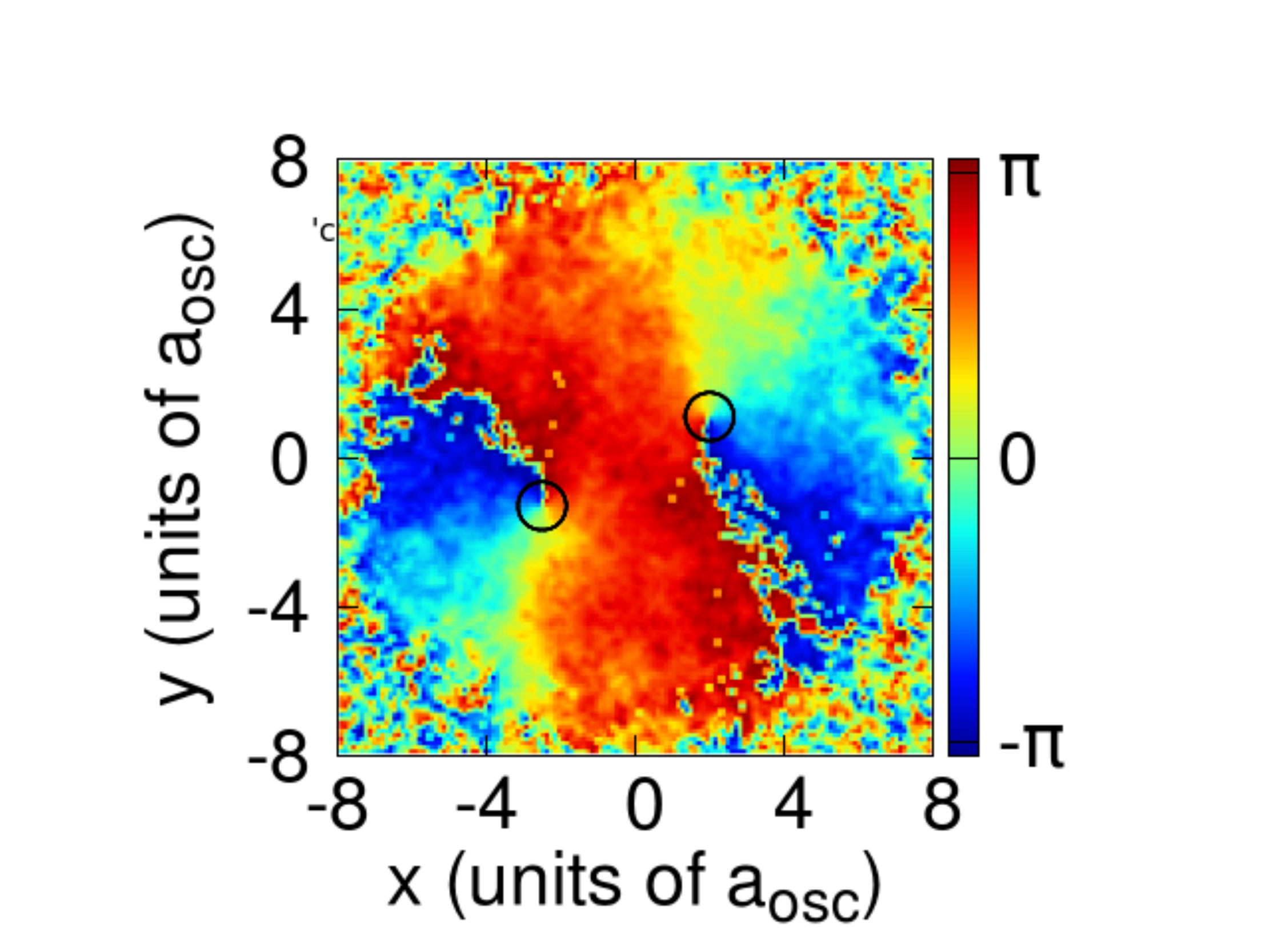}}\\
\end{tabular}
\caption{(Color online) The density (left column) and phase profiles 
(right column) of the condensate with two corotating vortices just before 
switching of the rotational frequency at $T = 30$ nK (upper panel) and $40$ nK
(lower panel). The locations of the vortices have been marked by two black 
circles in each image.}
\label{Fig-3}
\end{figure}
\end{center}

\begin{center}
\begin{figure}
\begin{tabular}{c}
\resizebox{!}{!}
{\includegraphics[trim = 0cm 0mm 0cm 0mm,clip, angle=0,width=9cm]
                 {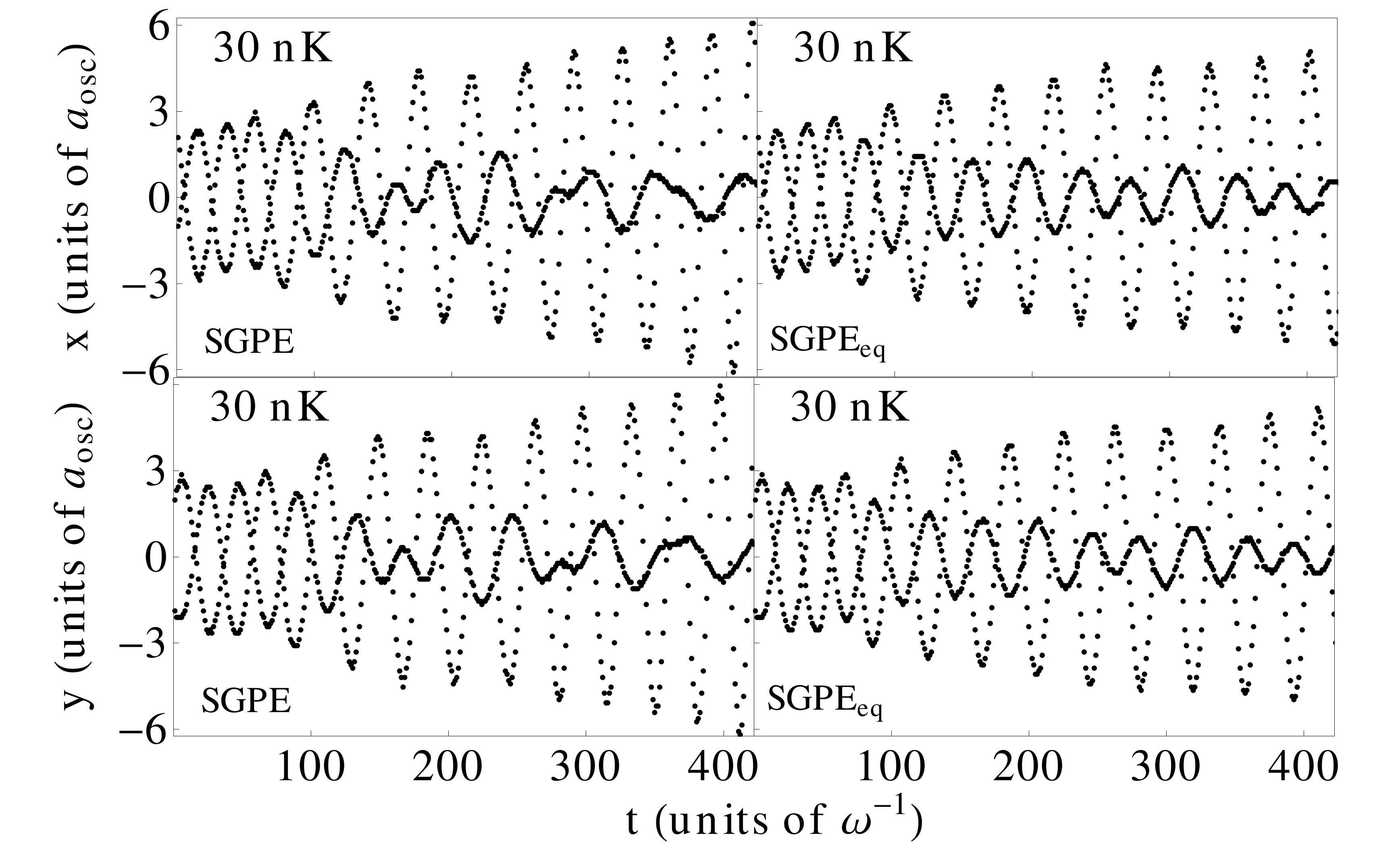}}\\
\resizebox{!}{!}
{\includegraphics[trim = 0cm 0mm 0cm 0mm,clip, angle=0,width=9cm]
                 {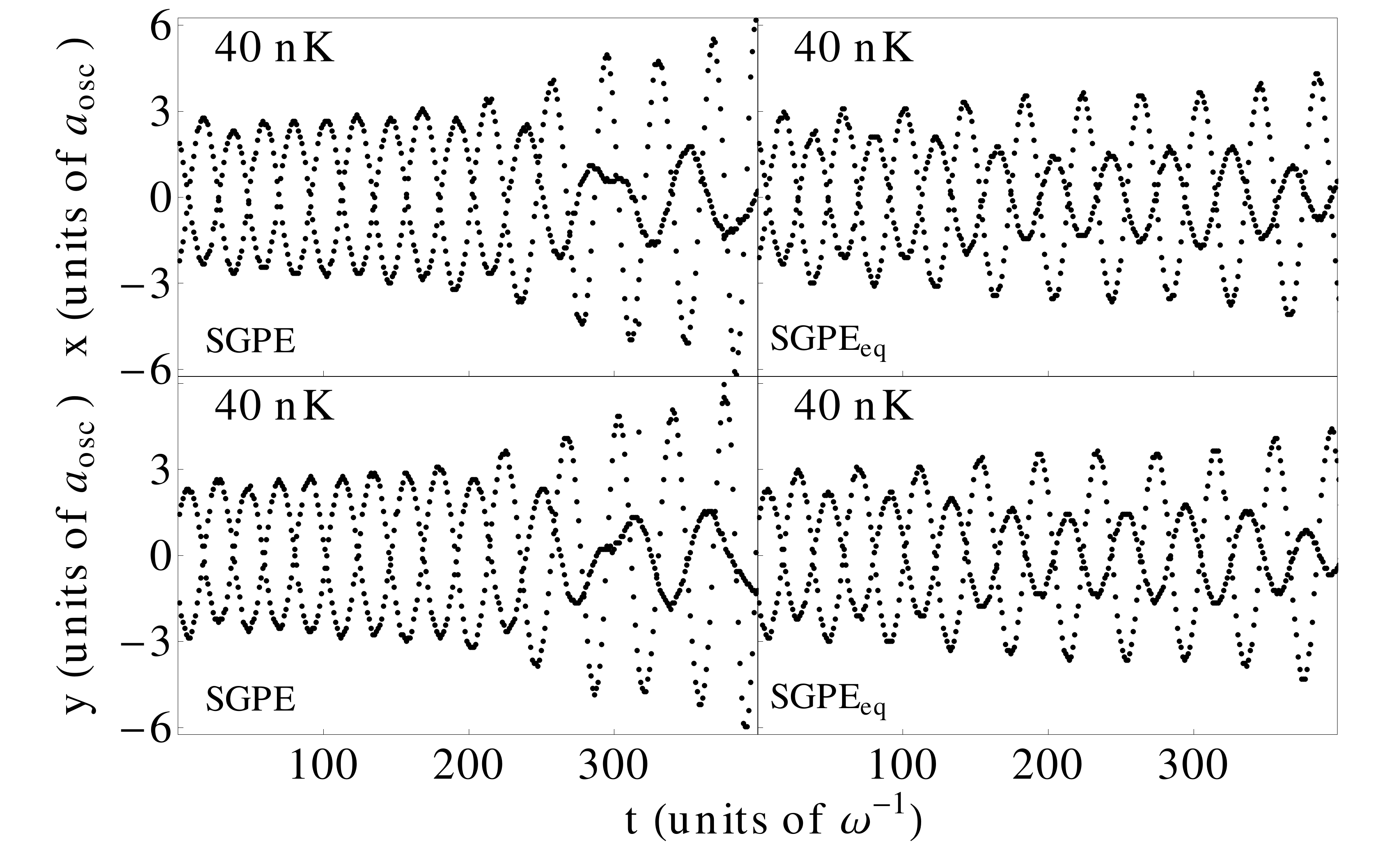}}\\
\end{tabular}
\caption{The $x$- and $y$-coordinates of the vortices as a function of time at 
$30$ nK and $40$ nK temperatures. It can be seen that there is an asymmetry in 
decay rates that increases with time.
}
\label{Fig-4}
\end{figure}
\end{center}

\begin{center}
\begin{figure}
\includegraphics[trim = 0cm 0mm 0cm 0mm,clip, angle=0,width=9cm]
                 {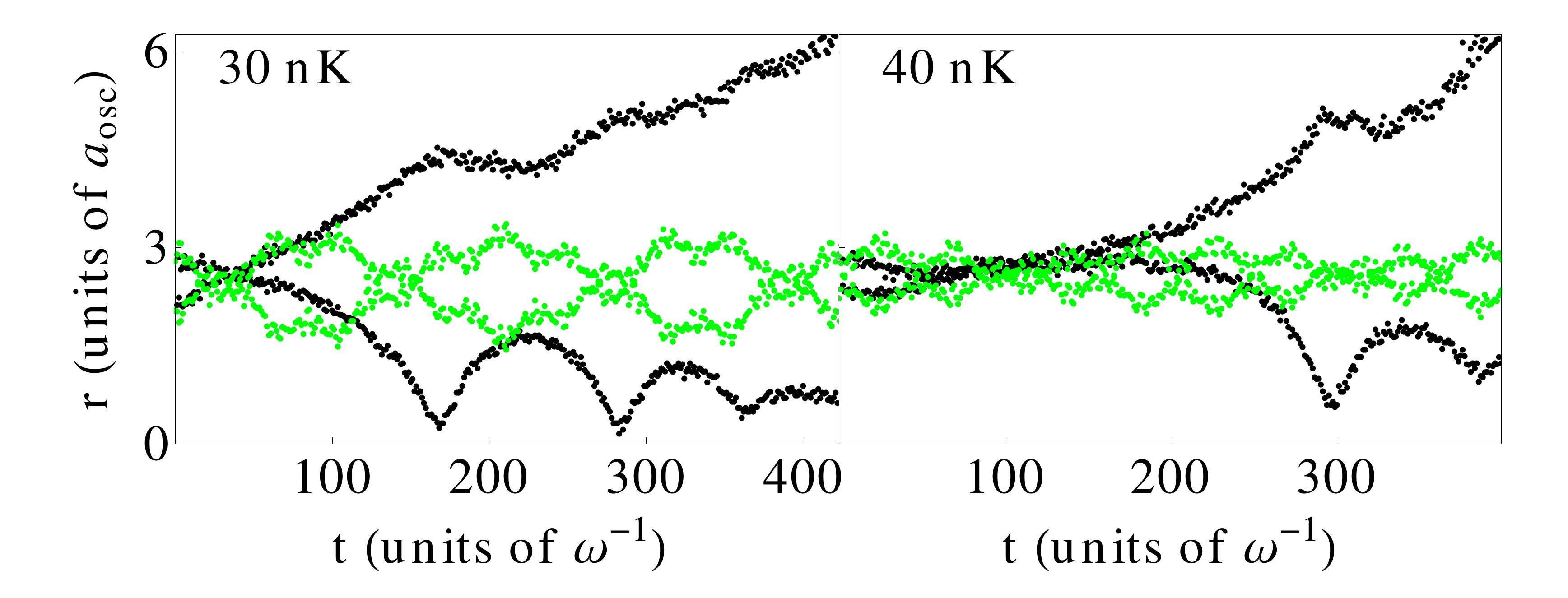}
\caption{The black dots are the radial coordinates of the vortices as a 
function of time at $30$ nK and $40$ nK temperatures obtained using SGPE 
method. The green dots are the radial coordinates of the vortices obtained
using GP equation with $N = 1.3\times10^5$ and $1.4\times10^5$ for the left
and right figures respectively.
}
\label{cv_dyn}
\end{figure}
\end{center}

Extending the method discussed in the previous section, we can also study the
dynamics of a corotating vortex pair at a finite temperature near the 
condensation temperature. We first generate a vortex pair by rotating the
condensate at finite temperature with a frequency $\Omega = 0.16~\omega_x$.
The density and phase profiles of the condensate before switching off the
rotational frequency at $30$ nK and $40$ nK temperatures are shown in
Fig.~\ref{Fig-3}. 
There are two vortices: one each in the first and third quadrant. We observe 
that one of the vortices, the one in the third quadrant at $30$ nK and the one 
in the first quadrant at $40$ nK, decays much faster than the other vortex as 
is evident from Fig.~\ref{Fig-4}. Again, the results obtained using the SGPE 
and SGPE$_{\rm eq}$ match during the initial period of evolution
(see Fig.~\ref{Fig-4}).
The faster decaying vortex suppresses the rate of decay of the other vortex, 
which moves towards the center of the trap. This is illustrated in 
Fig.~\ref{cv_dyn} (black dots) which shows the radial coordinates of the two 
vortices as a function of time obtained using SGPE.
Since there is a pair of interacting vortices, it is no longer true that the 
one which is closer to the edge of the trap will decay faster as would have 
been the case for isolated vortices. 
Which of the two vortices decays faster 
is determined by the complex interplay between three processes: (a) 
position-dependent vortex precession, (b) velocity field induced by one vortex 
at the location of the other, and (c) the random thermal fluctuations. 
Had the random thermal fluctuation been absent, as is the case for BEC at 
$T = 0~$K, the dynamics of the vortex pair would have been completely 
determined by the processes (a) and (b). 
In order to delineate the role of the 
these two processes, let us look at the $T = 0$ dynamics of two corotating 
vortices intially located at unequal radial distances. In order to do so, we 
first generate the vortex pair by imprinting the corresponding phase 
singularities at each (imaginary) time step while evolving Eq.~(\ref{Eq.1}) in
imaginary time, i.e., change
\begin{equation}
\Phi\rightarrow |\Phi| \exp \left\{i \left[\tan^{-1}\left(\frac{y-y_1}{x-x_1}
\right)+\tan^{-1}\left(\frac{y-y_2}{x-x_2}\right)\right]\right\}\nonumber
\end{equation}
after each time step. Here $(x_1,y_1)$ and $(x_2,y_2)$ are the locations of two
vortices and are chosen to be equal to the locations of vortices at $30$ nK and
$40$ nK just prior to switching off $\Omega$. The number of atoms is fixed to 
$1.3\times10^5$ and $1.4\times10^5$ for the case corresponding to $30$ nK and 
$40$ nK respectively. This solution is then evolved in real time again using 
Eq.~(\ref{Eq.1}). We observe that the two vortices rotate about the trap center 
with their respective radial coordinates oscillating with time. This is shown 
in Fig.~\ref{cv_dyn} with green dots. We find that due to these oscillations 
in the radial coordinates at $T = 0$ K, the two vortices alternatively end up 
being located at farther distance from the trap center as is evidenced by 
green dots in Fig.~\ref{cv_dyn}. The amplitude of these oscillations in the 
radial coordinates is more for more assymetry in thier initial locations.
Now, at finite temperatures, due to ocillating radial coordinates of the 
vortices during the intial stages of the evolution (the period during which 
the dynamics of the vortices will be qualitatively similar to $T=0$ dynamics), 
the random thermal fluctuations will result in one of the vortices decaying 
slightly faster than the other. While this vortex moves away from the trap 
center, the other vortex moves closer the trap center, reminiscent of their 
dynamics at $T=0$ K, due to the combined affect processes (a) and (b). As is 
the case at $T = 0$ K, the radial coordinates of the two vortices show an 
anticorrelated behaviour, i.e, when one is moving away from the trap center, 
the other is moving towards it. Hence, the combined effect of the processes 
(a), (b), and (c) results in a positive feedback whereby the rate of decay of 
the slower decaying vortex is suppressed and that of the faster decaying one 
is enhanced. The faster decaying vortex is removed quickly from the system, 
after which the second vortex will decay like a single isolated vortex.  

%%%%%%%%%%%%%%%%%%%%%%%%%%%%%%%%%%%%%%%%%%%%%%%%%%%%%%%%%%%%%%%%%%%%%%%%%%%%%%
%%%%%%%%%%%%%           Dynamics of a vortex dipole                %%%%%%%%%%%
%%%%%%%%%%%%%%%%%%%%%%%%%%%%%%%%%%%%%%%%%%%%%%%%%%%%%%%%%%%%%%%%%%%%%%%%%%%%%%

\subsection{Dynamics of a vortex dipole}
\begin{center}
\begin{figure}
\begin{tabular}{ccc}
\resizebox{!}{!}
{\includegraphics[trim = 2.1cm 2cm 1.6cm 2cm,clip, angle=0,width=2.8cm]
                 {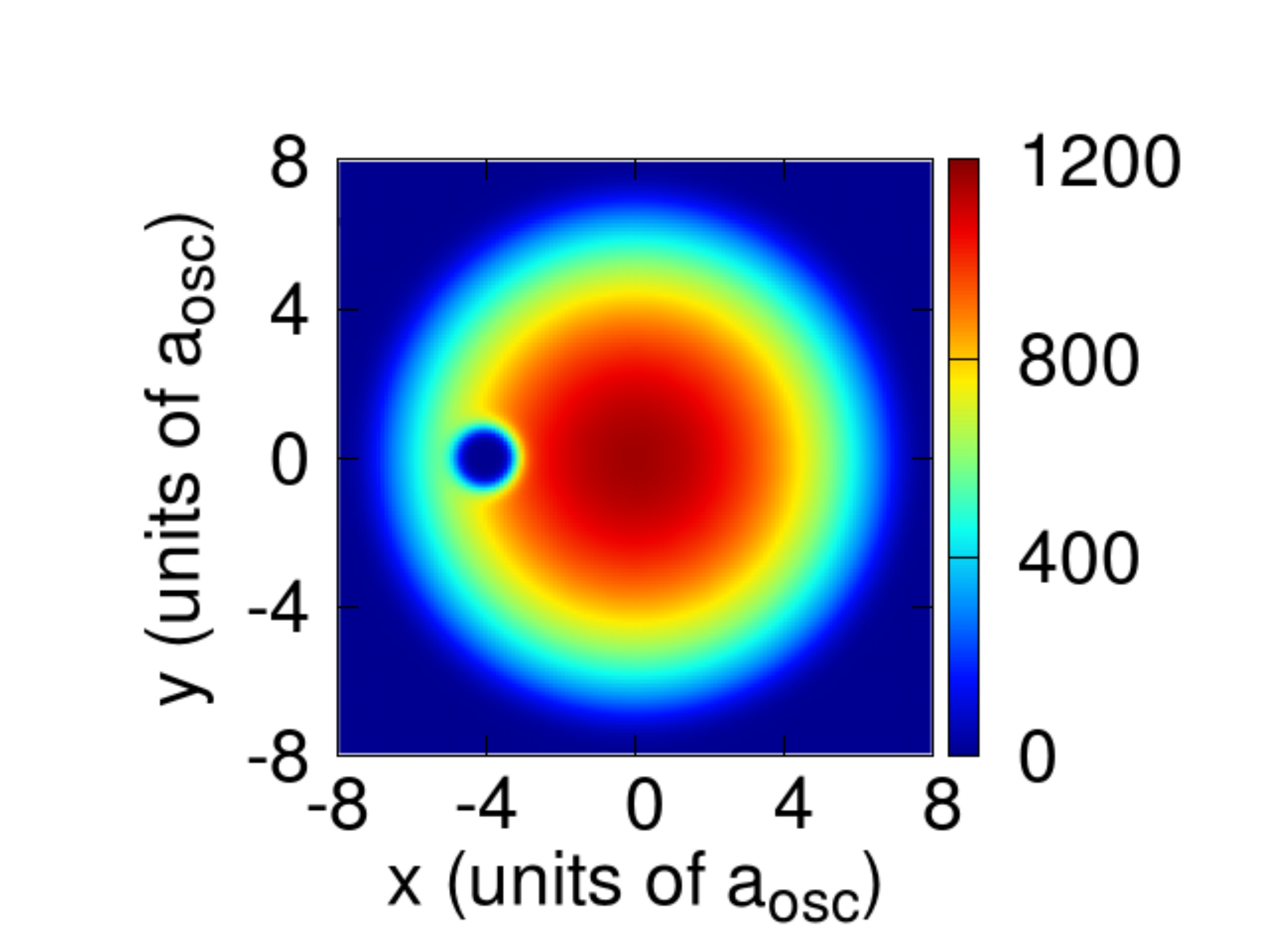}}
\resizebox{!}{!}
{\includegraphics[trim = 2.1cm 2cm 1.6cm 2cm,clip, angle=0,width=2.8cm]
                 {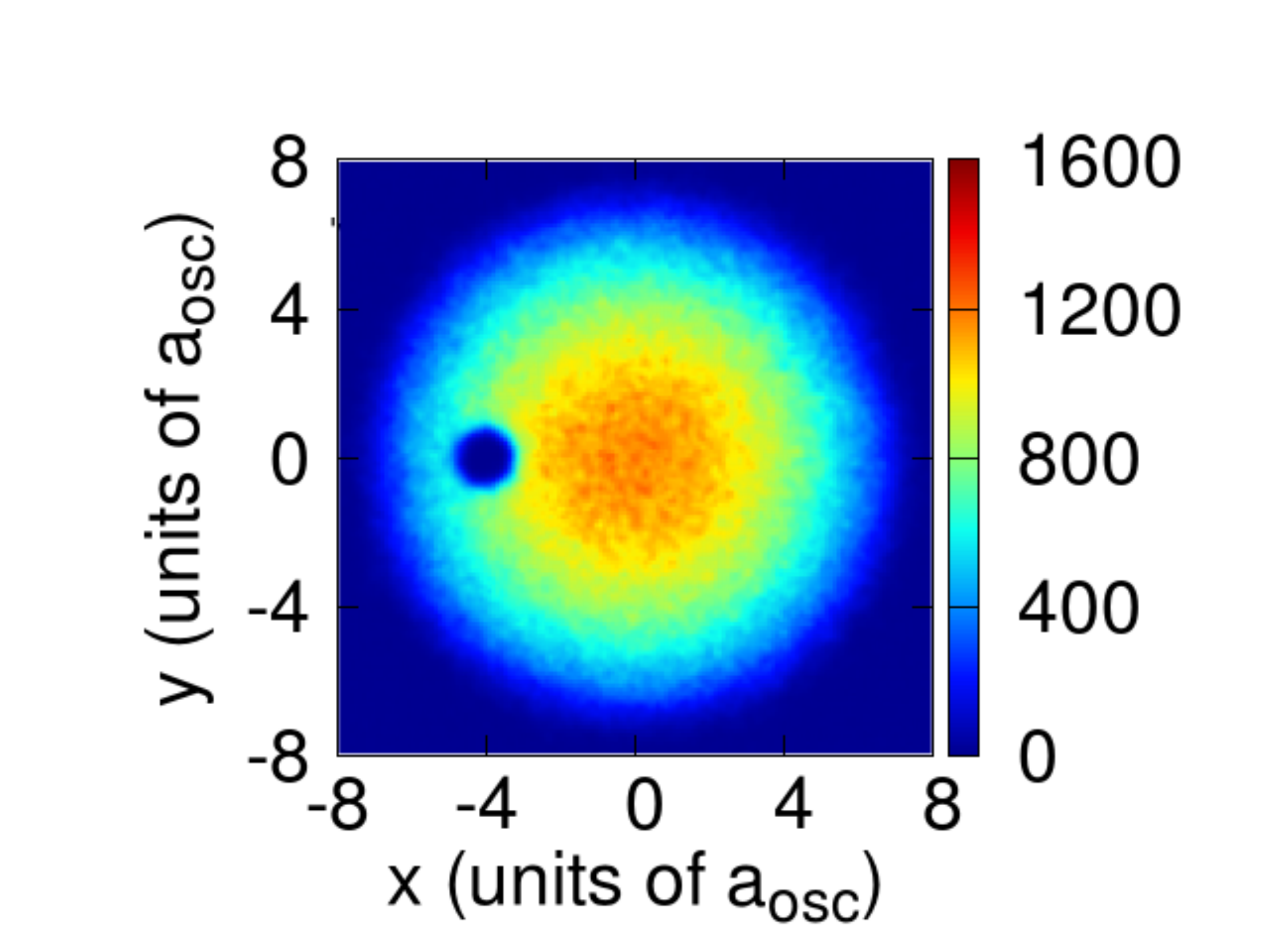}}
{\includegraphics[trim = 2.1cm 2cm 1.6cm 2cm,clip, angle=0,width=2.8cm]
                 {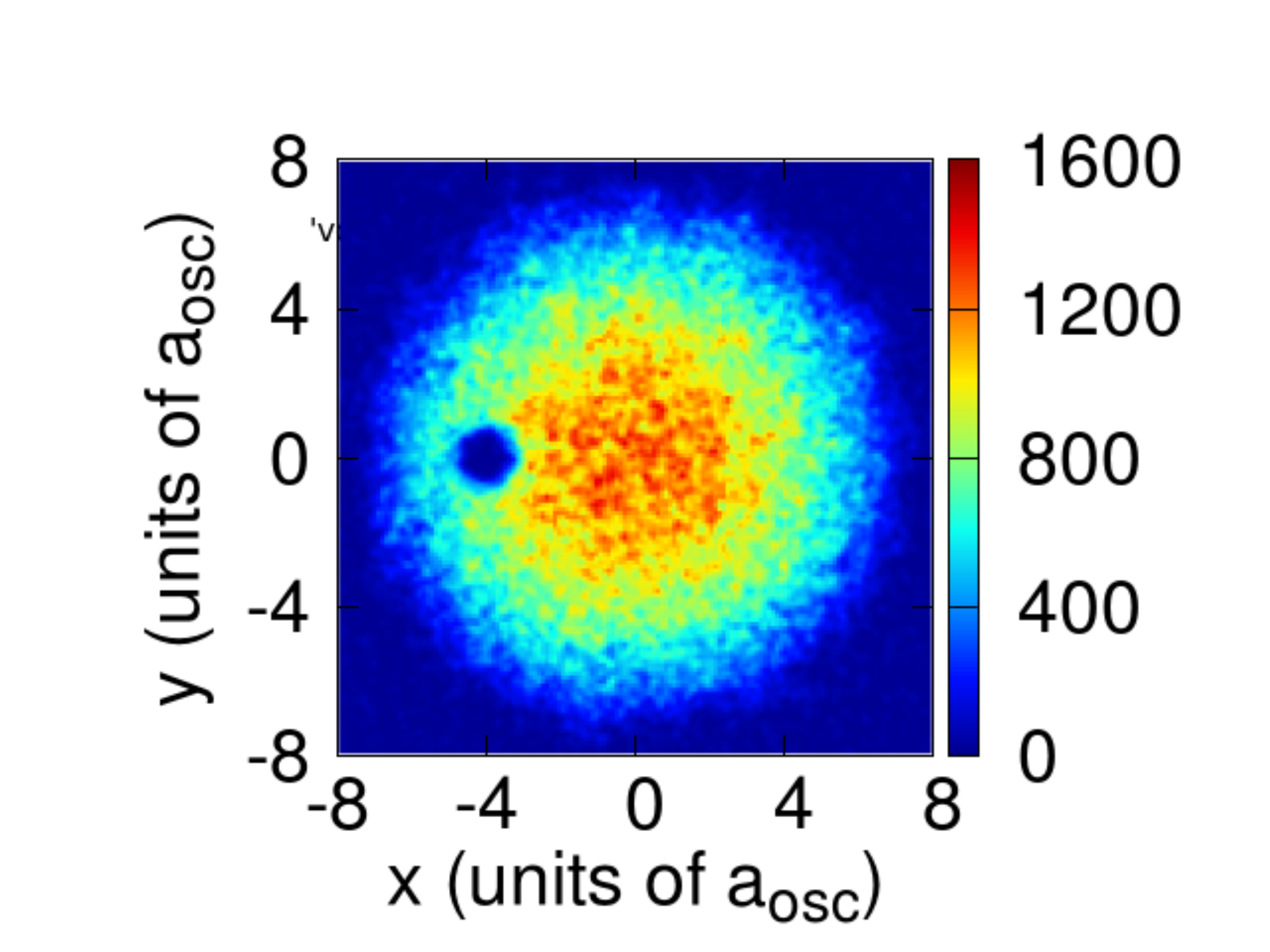}}\\
{\includegraphics[trim = 2.1cm 0cm 1.6cm 2cm,clip, angle=0,width=2.8cm]
                 {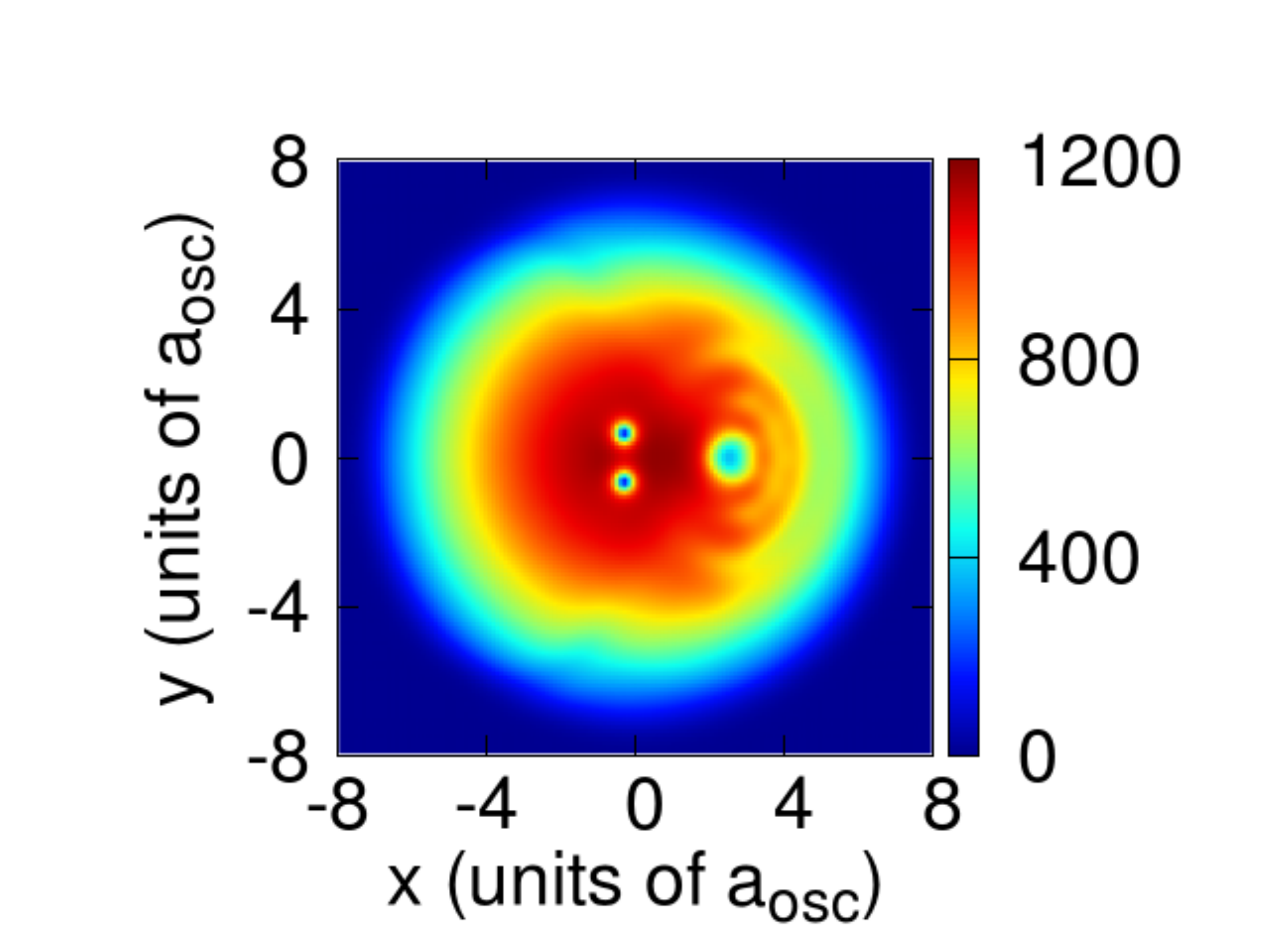}}
\resizebox{!}{!}
{\includegraphics[trim = 2.1cm 0cm 1.6cm 2cm,clip, angle=0,width=2.8cm]
                 {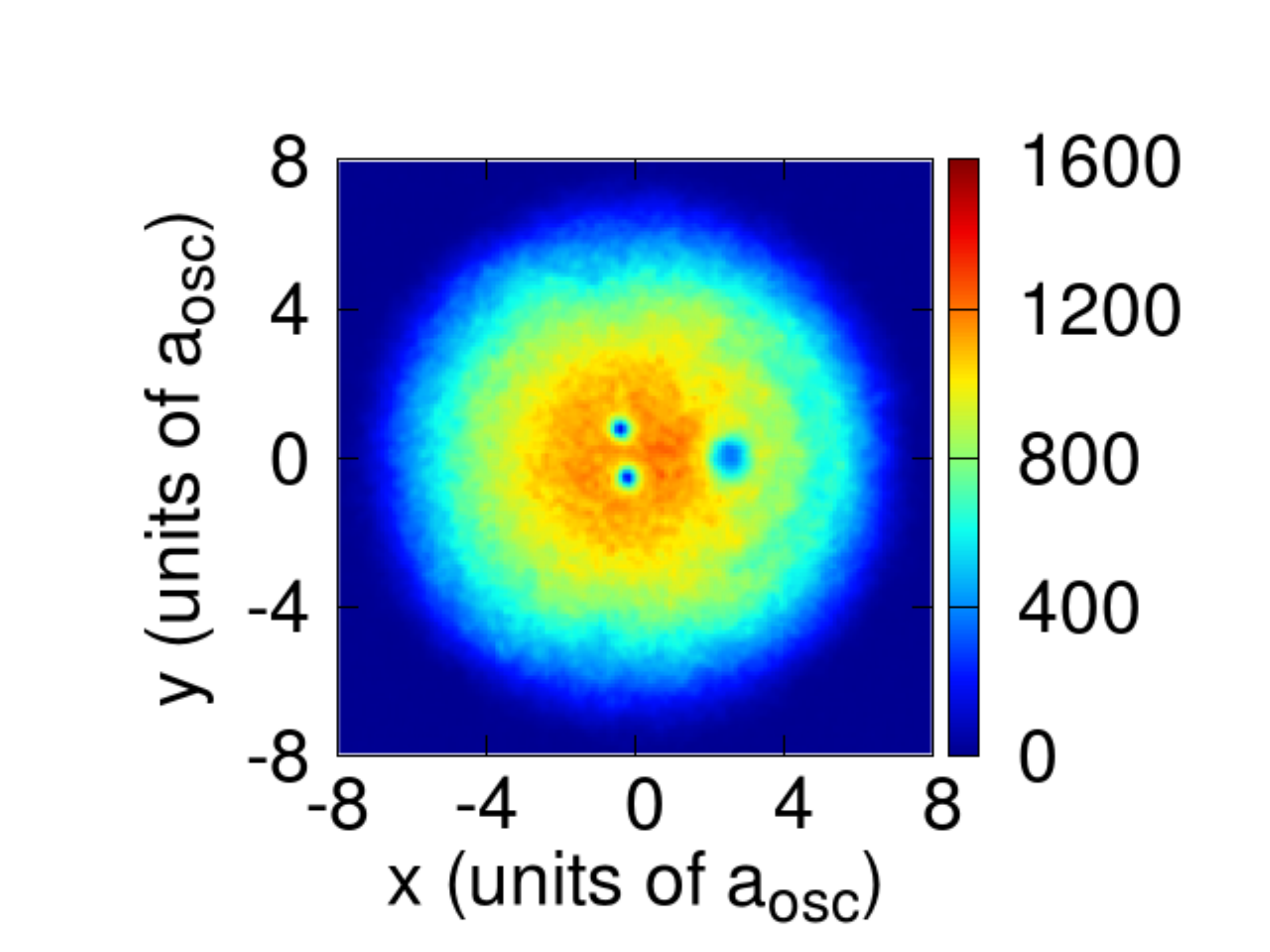}}
{\includegraphics[trim = 2.1cm 0cm 1.6cm 2cm,clip, angle=0,width=2.8cm]
                 {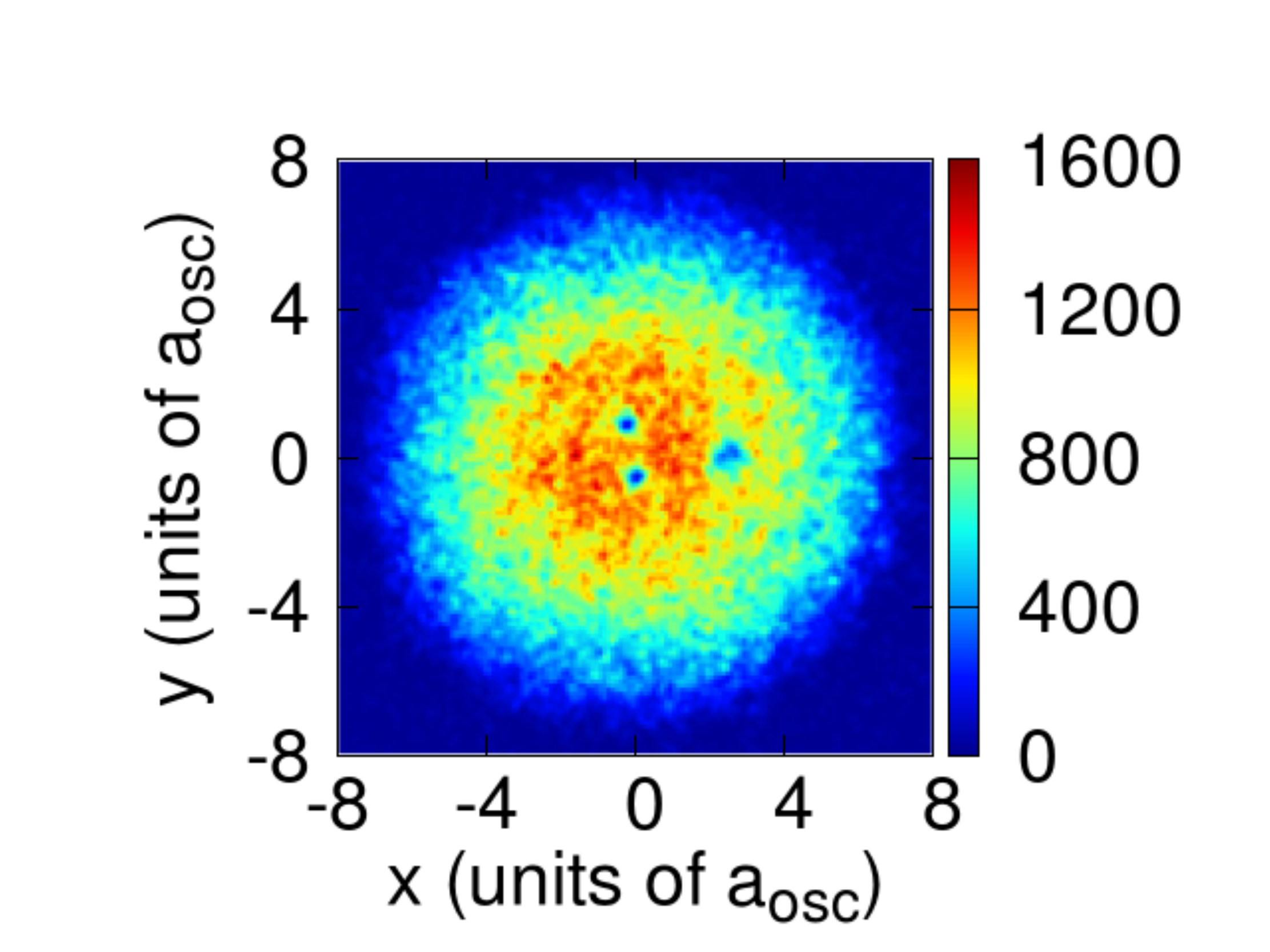}}
\end{tabular}
\caption{(Color online) The upper row shows the density distribution of the 
condensate with the obstacle potential at $-4~a_{\rm osc}$. The left, middle
and right images are at $T = 1$ fK, $1$ nK, and $10$ nK, respectively.
The bottom row shows the density distribution after
$t=3.659537~\omega^{-1}$.
}
\label{Fig-5}
\end{figure}
\end{center}

\begin{center}
\begin{figure}
\includegraphics[trim = 0cm 0mm 0cm 0mm,clip, angle=0,width=8cm]
                 {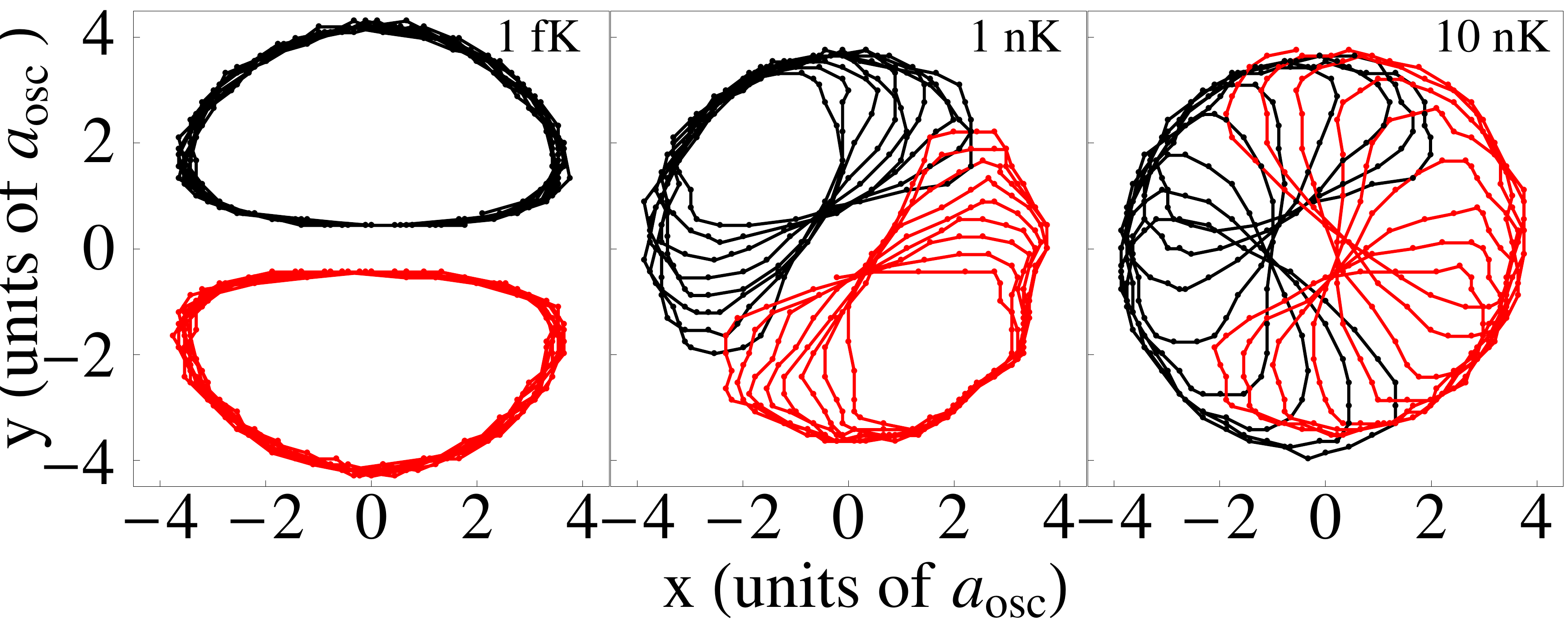}
\caption{(Color online) Trajectories of the vortex dipole at $1$ fK, $1$ nK, and
$10$ nK temperatures. The black and red colors correspond to vortex and 
antivortex respectively. The obstacle is moving at a speed of $380~\mu$m/s. 
The trajectories have been tracked for a period of $\approx 310~\omega^{-1}$.}
\label{Fig-7}
\end{figure}
\end{center}

\begin{center}
\begin{figure}
\includegraphics[trim = 0cm 0mm 0cm 0mm,clip, angle=0,width=9cm]
                 {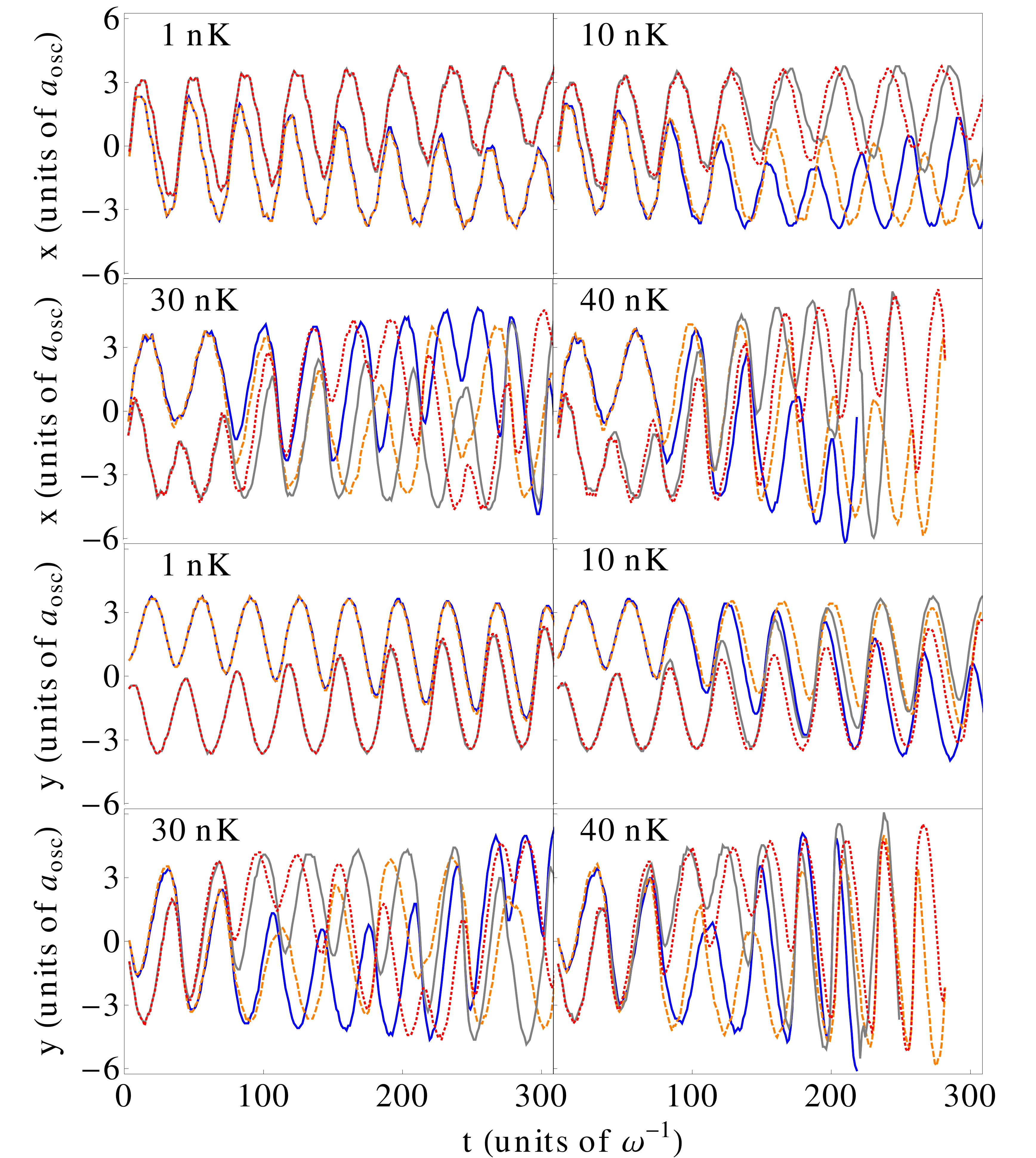}
\caption{(Color online) Vortex dipole dynamics at $1$, $10$, $30$ and $40$ nK.
The solid curves correspond to results obtained using the SGPE while the dotted and dashed curves correspond to the results obtained
using the SGPE$_{\rm eq}$. Solid-blue and dashed-orange curves
represent the dynamics of the vortex and solid-grey and dotted-red
represent the dynamics of antivortex.}
\label{Fig-8}
\end{figure}
\end{center}

Rotating the trapping potential can only lead to the formation of
co-rotating vortices. Hence, we need an alternative method to create the
vortex-antivortex pairs (vortex dipoles). It has been established both
theoretically and experimentally that when a superfluid moves past an impurity
faster than a critical speed, vortex dipoles are generated \cite{Frisch,Neely}.
In the present work, we create a vortex dipole in the BEC by moving a Gaussian 
obstacle potential across it above a critical speed. Recently, the method was 
also used by us to study the generation and stability of vortex-bright soliton 
dipoles in phase-separated binary condensates \cite{Gautam-1}. In our
simulations, we first generate the stationary solution without the obstacle
potential, i.e., only in the presence of a harmonic trapping potential. Then,
we slowly introduce the obstacle potential, keeping it fixed at a point. This
procedure ensures that no phase singularity is trapped by the obstacle
potential. In the present work, we consider $10^5$-$1.5\times10^5$
atoms of $^{87}$Rb (the exact number is a function of temperature) trapped in a
trapping potential with $\omega=\omega_x=\omega_y=2\pi\times10$ Hz and
$\omega_z=2\pi\times100$ Hz. In addition to this, there is a Gaussian obstacle 
potential
\begin{equation}
V_{\rm obs} = V_0 e^{\frac{-2\{(x-x_0(t))^2+y^2\}}{w_0^2}},
\end{equation}
where $(x_0,0)$ is the location of the obstacle potential, $V_0$ its amplitude,
and $w_0$ its width. In the present work
$x_0(0) = -4~a_{\rm osc}$, $V_0 = 100~\hbar\omega_x$ and $w_0=2.5~\mu$m.
The critical speed for these set of parameters at $1$ fK is $\sim 360~\mu$m/s.
The density profiles of the condensate obtained by using this
set of parameters at $1$ fK, $1$ nK and $10$ nK are shown in the upper
row of Fig.~\ref{Fig-5}.
In order to generate the vortex dipole, the obstacle is moved
along the $x$-axis at an optimum speed. As the obstacle is moved, its
strength is continuously decreased so that the obstacle potential vanishes at
$x=4~a_{\rm osc}$. We observe that at sufficiently low temperatures,
the vortices are generated symmetrically about the $x-$axis, but not
at higher temperatures (see images in the lower row of
Fig.~\ref{Fig-5}). This asymmetry in the generation of
vortex-antivortex pairs was also pointed by us in an earlier
work \cite{Prabhakar}. The asymmetry in location of the vortex and antivortex
has also been observed experimentally as is evident from the inset of Fig.~2
in Ref. \cite{Neely}. Due to this asymmetry, the trajectory of a vortex
dipole at higher temperatures is qualitatively different from near zero
temperatures (say of the order of a few fK) as is shown in
Fig~.\ref{Fig-7}.
We have employed both the SGPE$_{\rm eq}$ and SGPE to study vortex dipole
generation and its dynamics at $1$ nK, $10$ nK, $30$ nK and $40$ nK.
As in the case of a single vortex, we find that the results
obtained with two methods are in good agreement during the initial period
of evolution at all temperatures as shown in Fig.~\ref{Fig-8}.
Also evident from the Fig.~\ref{Fig-8} is that as the temperature
is increased the vortex dipole decays faster. In fact at $40$ nK, the vortex 
dipole has already decayed before
$t= 300~\omega^{-1}$ as shown in Fig.~\ref{Fig-8}.
We find that at lower temperatures of $1$ nK, $10$ nK, and $30$ nK, the 
component vortices of the dipole decay at approximately same rate. At $40$ nK 
temperature, the component vortices decay at slightly different rate. As a 
result of it, the vortex (represented by solid-blue curve in Fig.~\ref{Fig-8}) 
is ejected out of the condensate earlier than the antivortex (represented
by solid-grey curve in Fig.~\ref{Fig-8}). The roughly similar rate of decay of 
the vortex and antivortex is due to the attractive interaction between them, 
in contrast to the case of the two co-rotating vortices, where the interaction 
is repulsive. In the case of the dipole, a negative feedback that tends to 
oppose any asymmetry in the decay rate between the vortex and antivortex 
since the elimination of only one tends to increase the energy of the system. 
Thus, both tend to decay at roughly the same rate with the asymmetry becoming 
more pronounced at higher temperature as the strength of thermal fluctuations 
(responsible for the asymmetry) increases.

\begin{center}
\begin{figure}[!h]
\includegraphics[trim = 0cm 0mm 0cm 0mm,clip, angle=0,width=6cm]
                 {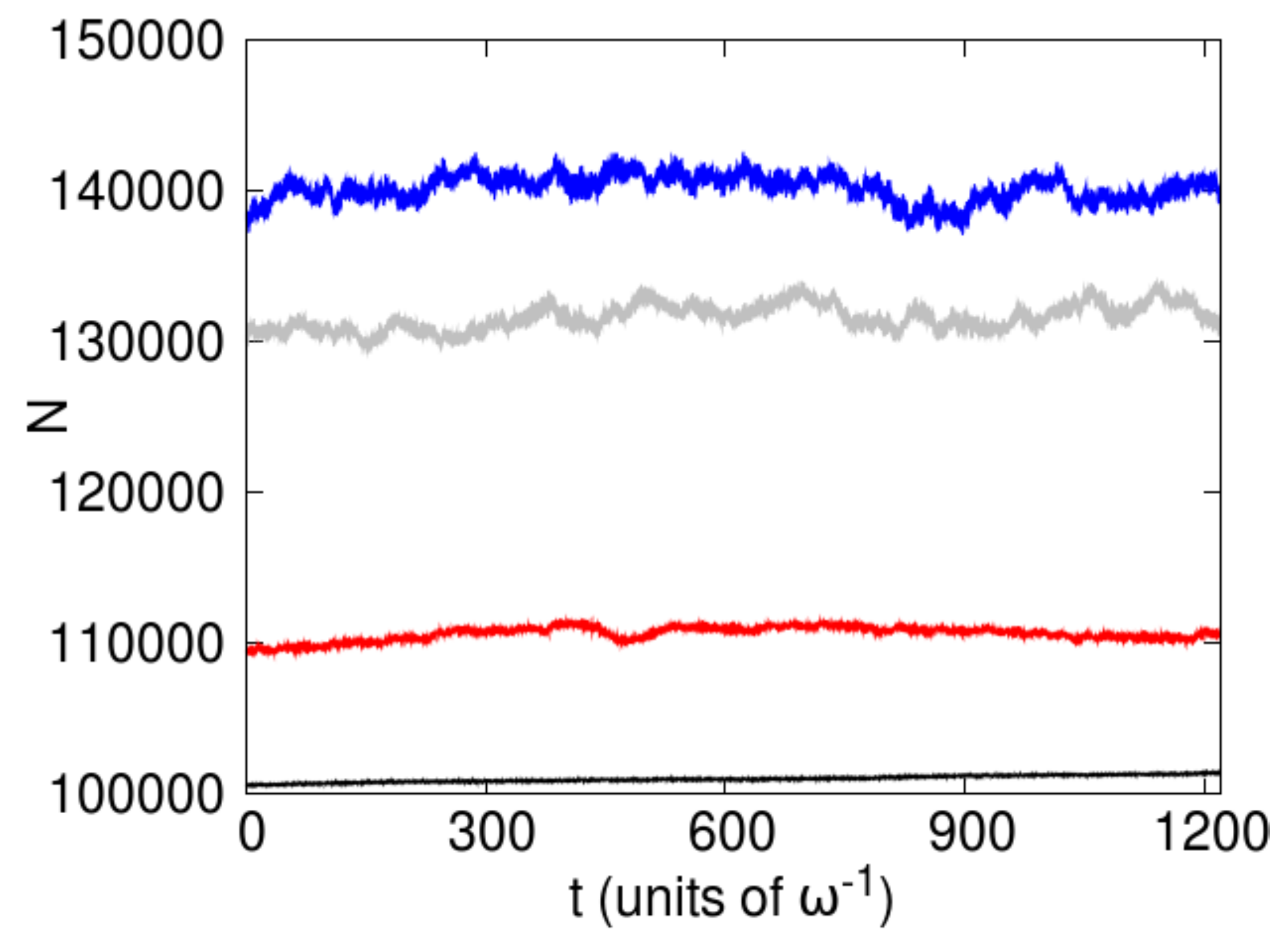}
\caption{(Color online) The plot shows the number of atoms as function of time,
obtained using SGPE, after the obstacle starts moving. The black, red, grey 
and blue curves correspond to $1$, $10$, $30$, and $40$ nK respectively. In the
SGPE$_{\rm eq}$, the number remains fixed to the value at $t=0$ at all
temperatures.}
\label{num_atoms2}
\end{figure}
\end{center}

The variation in the number of atoms after the obstacle starts moving
is shown in Fig.~\ref{num_atoms2}.

%%%%%%%%%%%%%%%%%%%%%%%%%%%%%%%%%%%%%%%%%%%%%%%%%%%%%%%%%%%%%%%%%%%%%%%%%%%%%%%
%%%%%%%%%%%%             Summary of results                        %%%%%%%%%%%%
%%%%%%%%%%%%%%%%%%%%%%%%%%%%%%%%%%%%%%%%%%%%%%%%%%%%%%%%%%%%%%%%%%%%%%%%%%%%%%%

\section{Summary and conclusions}
\label{V}
We have studied the dynamics of a single and a pair of vortices in quasi
two-dimensional systems at finite temperatures using the SGPE and
SGPE$_{\rm eq}$ methods. These methods describe BECs well in the regime where 
thermal fluctuations are significantly high. We find that like a single vortex,
a pair of vortices tends to decay as the system evolves. The rate of decay 
depends upon the temperature and the initial location of the constituent 
vortices. We find that on shorter time scales the dynamics of the system is 
basically governed by the multi-mode order parameter, which represents a few 
highly occupied low lying modes. For this reason, the results of the 
SGPE$_{\rm eq}$, which governs the intrinsic evolution of the multimode order 
parameter, agree with those obtained from the SGPE during this period. The 
higher modes affect the dynamics of the system on longer time scales. For a 
pair of vortices, the two vortices are not symmetrically generated with respect
to each other. In the case of corotating vortices, this initial asymmetry in 
their generation, coupled with random thermal fluctuations and a positive 
feedback resulting from the repulsive interaction between them leads to 
different decay rates for the two vortices. In contrast, the component 
vortices of a dipole decay at approximately same rate due to a negative 
feedback arising from the attractive interaction between them. At higher 
temperatures where thermal fluctuations are stronger, the asymmetry in the 
decay rates gets more pronounced.

%%%%%%%%%%%%%%%%%%%%%%%%%%%%%%%%%%%%%%%%%%%%%%%%%%%%%%%%%%%%%%%%%%%%%%%%%%%%%%%
%%%%%%%%%%%%             Acknowledgements                          %%%%%%%%%%%%
%%%%%%%%%%%%%%%%%%%%%%%%%%%%%%%%%%%%%%%%%%%%%%%%%%%%%%%%%%%%%%%%%%%%%%%%%%%%%%%

\begin{acknowledgements}
SG and SM would like to thank the Department of Science and Technology, 
Government of India for support. 
\end{acknowledgements}

%%%%%%%%%%%%%%%%%%%%%%%%%%%%%%%%%%%%%%%%%%%%%%%%%%%%%%%%%%%%%%%%%%%%%%%%%%%%%%%
%%%%                        Bibliography                                  %%%%%
%%%%%%%%%%%%%%%%%%%%%%%%%%%%%%%%%%%%%%%%%%%%%%%%%%%%%%%%%%%%%%%%%%%%%%%%%%%%%%%

\end{document}